\documentclass[preprint]{elsarticle}
\usepackage[bookmarks=false]{hyperref}
\usepackage{graphicx}
\usepackage{float}
\usepackage[export]{adjustbox}
\usepackage{algorithm}
\usepackage[noend]{algpseudocode}
\usepackage{footnote}
\usepackage{fullpage}
\usepackage{amssymb}
\usepackage{bbm}
\usepackage{microtype}
\usepackage[textsize=scriptsize, textwidth = 2.2cm]{todonotes}
\usepackage{amsmath}
\usepackage{lipsum}
\usepackage{scalerel}

\usepackage{amsfonts,bm}
\def\vw{{\bm{w}}}
\def\vg{{\bm{g}}}

\def\sX{{\mathbb{X}}}
\def\sB{{\mathbb{B}}}
\def\sT{{\mathbb{T}}}
\newcommand{\E}{\mathbb{E}}
\newcommand{\Var}{\mathrm{Var}}
\DeclareMathOperator*{\argmax}{arg\,max}
\usepackage{mleftright}
\newcommand{\adasync}{\textsc{AdaSync}}
\newcommand{\DynamicBackUp}{\textsc{DBW}}
\newcommand{\bdbw}{\textsc{B-DBW}}

\newcommand{\V}{\mathbb{V}}

\usepackage[labelformat=simple]{subcaption}

\date{}
\DeclareMathOperator*{\minimize}{minimize}

\begin{document}

\begin{frontmatter}

\title{Dynamic Backup Workers for Parallel Machine Learning}

\author{Chuan Xu}
\ead{chuan.xu@inria.fr}
\author{Giovanni Neglia}
\ead{giovanni.neglia@inria.fr}
\author{Nicola Sebastianelli}
\ead{nicola.sebastianelli@inria.fr}
\address{Inria, Universit\'e C\^ote d'Azur, Sophia Antipolis, France}

\begin{abstract}
The most popular framework for distributed training of machine learning models is the (synchronous) parameter server (PS). This paradigm consists of $n$ workers, which  iteratively compute updates of the model parameters, and a stateful PS, which waits and aggregates all updates to generate a new estimate of model parameters and sends it back to the workers for a new iteration. 
Transient computation slowdowns or transmission delays can intolerably lengthen the time of each iteration. An efficient way to mitigate this problem is to let the PS wait only for the fastest $n-b$ updates, before generating the new parameters. The slowest $b$ workers are called \emph{backup workers}. 
The correct choice of the number $b$ of backup workers depends on the cluster configuration and workload, but also (as we show in this paper) on the hyper-parameters of the learning algorithm and the current stage of the training. 
We propose \DynamicBackUp, an algorithm that dynamically decides the number of backup workers during the training process to maximize the convergence speed at each iteration. Our experiments show that \DynamicBackUp{} 1) removes the necessity to tune $b$ by preliminary time-consuming experiments, and 2) makes the training up to a factor $3$ faster than the optimal static configuration. 
\end{abstract}

\end{frontmatter}
%\nolinenumbers

\section{Introduction}
Already in 2014, state-of-the-art machine learning models counted hundreds of billions of parameters and required processing hundreds of terabytes through thousands of cores~\cite{canini14}. As models and datasets keep becoming larger, the need for efficient distributed  solutions  becomes even more urgent. 
These distributed systems are different from those used for traditional applications like transaction processing or data analytics, because of statistical and algorithmic characteristics unique to ML programs, like error tolerance, structural dependencies, and non-uniform convergence of parameters~\cite{xing16}.
Currently, their operation requires a number of ad-hoc choices and time-consuming tuning through trial and error, e.g.,~to decide how to distribute ML programs over a cluster or how to bridge ML computation with inter-machine communication. For this reason, significant research effort (also from the networking community~\cite{Harlap,wang19,shi19,bao19, CChen, Shi, 9155448})
is devoted to design adaptive algorithms for a more effective use of computing resources for ML training.

%Significant engineering efforts are required to run these distributed systems.
%The example in~\cite{young17} is typical of a large class of iterative  ML distributed algorithms. Such algorithms begin with a guess of an optimal vector of parameters and proceed through multiple iterations over the input data to improve the solution. The process evolves in a data-parallel manner: input data is divided among worker threads, each of which iterates over its data subset and determines solution adjustments based on its local view of the latest parameter values.  Solution adjustments are then exchanged among workers. 
%This operation appears to fit general-purpose computing frameworks like Apache Spark, which was designed to extend MapReduce and other data parallel abstractions to the iterative setting.

For distributed ML training, there are two popular frameworks, the parameter server (PS)~\cite{Li:2014} and AllReduce (AR)~\cite{DBLP:journals/jpdc/PatarasukY09, tree-basedAR}. In PS, a stateful parameter server maintains the current version of the model parameters and broadcasts them to the workers (computing units e.g.,~GPUs). Every worker then computes ``delta'' updates of the parameters, e.g.,~through a gradient descent step. These updates are then aggregated by the PS in a synchronized way and combined with its current state to produce a new estimate of the optimal parameter vector. As the server may become a communication bottleneck, aggregation can be implemented in a distributed way through an AllReduce collective operation~\cite{mpiall}.
For example, in Ring-AllReduce~\cite{baidu} with $n$ workers, $2(n-1)$ synchronized communications are required with $\mathcal O(1)$ data transmitted per worker.
However, both the PS and AR are sensitive to  \emph{stragglers}~\cite{dean2013tail,ananthanarayanan13,karakus17,LiKAS18, DBLP:journals/corr/abs-1902-01981}, i.e., ``\emph{workers that are \textbf{randomly} slowed down due to resource contention, background OS activities,  garbage collection, and (for ML tasks) stopping criteria calculations}"~\cite{Harlap}.

To mitigate the stragglers problem, coding techniques have been proposed 
%(including to code the computation or to code the transmission of the gradients) 
both for PS~\cite{ pmlr-v70-tandon17a, lee2017speeding, DBLP:conf/isit/HalbawiRSH18, pmlr-v89-yu19b,reisizadeh2019coded, ozfatura2020straggler, LiKAS18} and AR~\cite{reisizadeh2019tree, DBLP:journals/corr/abs-1902-01981} frameworks. The main idea behind is that each worker performs some additional computation and codes its update in an opportune way, so that only a subset of the tasks is needed to recover the full information and to proceed to the next iteration. Hence, the system does not need to wait for the stragglers.
Coding techniques are particularly helpful when data distribution across workers is heterogeneous~\cite{dutta} as it happens in federated learning~\cite{mcmahan2017communication}.
In a cluster, all workers have access to the whole dataset or to a random sample of it, hence the advantage of coding is significantly reduced, 
and when computation time is larger than communication time, coding is even less beneficial~\cite{pmlr-v70-tandon17a}. In these settings, the additional overhead introduced by coding techniques may not be justified.

Alternative approaches to deal with stragglers are based on load-aware and interference-aware resource scheduling to monitor and avoid  stragglers~\cite{peng2018optimus, bao19}. These techniques are effective only if stragglers are \emph{persistent}, i.e., the same workers are slow over a relatively long time period, but straggler effects often occur over short timescale.

Another possibility is to relax the full  synchronization requirement avoiding to collect information from all workers before computing the new model parameters. One solution is to let the PS operate asynchronously, 
updating the parameter vector as soon as  it receives the result of a single worker~\cite{cipar2013solving, Dutta2018}.
While this approach increases system throughput (parameter updates per time unit),  workers operate in general on stale versions of the parameter vector slowing and, in some cases, even preventing convergence to the optimal model~\cite{DaiZDZX19}. Another solution is to apply decentralized learning methods, where there is no central server, but workers communicate only with their neighbours on an opportune communication graph~\cite{nedic-consensus,duchi12,lian17,elgabli2020gadmm}. When the graph is sparse and the stragglers behave in a non-persistent way, such methods work well enjoying high system throughput and guaranteed convergence~\cite{neglia19infocom,NegliaXTC20,marfoq20neurips}.
However, persistent stragglers can still slow down dramatically the throughput performance.

In the PS architecture, a simple solution to mitigate the effect of stragglers without  jeopardizing convergence, is to rely on backup workers~\cite{Jianmin2016, dutta}: instead of waiting for the updates from all workers (say it $n$), the PS waits for the fastest $k$ out of $n$ updates to proceed to the next iteration. The remaining $b \triangleq n-k$ workers are called backup workers.\footnote{
    We stick to the name used in the original paper~\cite{Jianmin2016}, even if it is someway misleading, because backup workers do not replace other workers when needed.
    %(as the name suggests). 
    In fact all workers operate identically, and who are the backup workers change from one iteration to the other depending on their execution times at that specific iteration.
} 
Experiments on Google cluster with $n=100$ workers show that~a few backup workers ($4$--$6$) can reduce the training time by 30\% in comparison to the synchronous PS and by 20\% in comparison to the asynchronous PS~\cite{Jianmin2016}. 

The number of backup workers $b$ has a double effect on the convergence speed. 
The larger $b$ is, the faster each iteration is, because the PS needs to wait less inputs from the workers. 
At the same time, the PS aggregates less information, so the model update is noisier and more iterations are required to converge.
Currently, 
the number of backup workers is configured manually through some experiments, before the actual training process starts.
However, the optimal static setting is highly sensitive to the cluster configuration (e.g.,~GPU performances and their connectivity) as well as to its instantaneous workload.
Both cluster configuration and workload
may be unknown to the users (specially in a virtualized cloud setting) and may change as new jobs arrive/depart from the cluster. Moreover, in this paper we show that the choice of the 
number of backup workers 1) should depend also on hyper-parameters\footnote{
    An hyper-parameter is a parameter of the learning algorithm (and not of the model), but it can still influence the final model learned. 
} 
like the batch size, and 2) should change during the training itself (!) as the loss function approaches a (local) minimum. 
Therefore, the static configuration 
%proposed in~\cite{Jianmin2016} for 
of backup workers does not only require time-consuming experiments, but is particularly inefficient and fragile.

In this paper we propose the algorithm \DynamicBackUp{} (for Dynamic Backup Workers) that dynamically adapts the number of backup workers during the training process without prior knowledge about the cluster or the optimization problem.
 Our algorithm identifies the sweet spot between  the two contrasting effects of $b$
 (reducing the duration of an iteration and increasing the number of iterations for convergence), by maximizing at each iteration the decrease of the loss function \emph{per time unit}.

This paper extends our conference submission~\cite{DBLP:conf/networking/XuNS20} and is organized as follows.  Sect.~\ref{s:background}  provides relevant background and introduces the notation. Sect.~\ref{s:algorithm} illustrates the different components of our algorithm \DynamicBackUp{} with their respective preliminary assessments.
%and presents some preliminary assessments on their respective effects. 
\DynamicBackUp{} is then evaluated on ML problems in Sect.~\ref{s:experiments}.
The results show that \DynamicBackUp{} is robust to different cluster environments and different hyper-parameters' settings.
\DynamicBackUp{} does not only remove the necessity to configure an additional parameter ($b$) through costly experiments, but also reduce the training time by a factor as large as 3 in comparison to the best static configuration.
Sect.~\ref{s:conclusions} concludes the paper and discusses future research directions.
The code of our implementation is available online~\cite{github_dbw}.

\section{Background and notation}
\label{s:background}
Given a dataset $\sX=\{x_l, l=1, \dots S\}$, the training of ML models usually requires to find a parameter vector $\vw \in \mathbb R^d$ %(approximately) 
minimizing a loss function: %  over the dataset:

\begin{equation}
\label{e:opt_probl}
 \minimize_{\vw \in \mathbb R^d}{}  \;\; F(\vw)\triangleq \frac{1}{S} \sum_{l=1}^S f(x_l,\vw),
\end{equation}

where $f(x_l,\vw)$ is the loss of the model $\vw$ on the datapoint $x_l$. For example, in supervised learning, each point of the dataset is a pair $x_l= (\chi_l, y_l)$, consisting of an input object $\chi_l$ and a desired output value $y_l$. In the standard linear regression
method $\chi_l \in \mathbb R^d$, $y_l \in \mathbb R$, the input-output function is a linear one ($\hat y_l = \chi_l ^\intercal \vw$) and the loss function is the mean squared error $(\chi_l ^\intercal \vw -y_l)^2$.
More complex models like neural networks look for an input-output mapping in a much larger and more flexible family of functions, but they are trained solving an optimization problem like~\eqref{e:opt_probl}.

The standard way to solve~Problem~\ref{e:opt_probl} is to use an iterative gradient method. Let $n$ be the number of workers (e.g.,~GPUs) available. In a synchronous setting without backup workers, at each iteration $t$ the PS sends the current estimate of the parameter vector $\vw_t$ to all workers. Each worker computes then a stochastic gradient on a random mini-batch of size $B$ ($\le S$) drawn from its local dataset. 
%For simplicity, 
We assume each worker has access to the complete dataset $\sX$ as it is resonable in the cluster setting that we consider.
%, but the algorithm works with minor modifications also when each worker has a different local dataset. 
Each worker sends the stochastic gradient back to the PS. We denote by $\vg_{i,t}$ the $i$-th worker gradient received by the PS at iteration $t$, i.e.,

\begin{equation}
\vg_{i,t} = \frac{1}{B} \sum_{x \in \sB_i} \nabla f(x, \vw_t),   \label{e:gradient} 
\end{equation}

and $\sB_i\subseteq \sX$ is the random minibatch of size $B$ on which the gradient has been computed.
Once $n$ gradients are received, the PS computes the average gradient %$\vg_t$

\[\vg_t = \frac{1}{n} \sum_{i=1}^{n} \vg_{i,t},\]

and updates the parameter vector as follows:

\begin{equation}
\label{e:update}
\vw_{t+1} = \vw_t - \eta \vg_t,
\end{equation}

where $\eta >0 $ is called the learning rate.

When $b$ backup workers are used \cite{Jianmin2016}, the PS only waits for the first $k=n-b$ gradients and then evaluates the average gradient as 

\begin{equation}
\label{e:grad_bw}
\vg_t = \frac{1}{k} \sum_{i=1}^{k} \vg_{i,t}.    
\end{equation}

In our dynamic algorithm (Sect.~\ref{s:algorithm}), the value of $k$ is no longer static but changes in an adaptive manner from one iteration to the other, ensuring faster convergence speed. 
We denote by $k_t$    the number of gradients of $\vw_t$  the PS needs to wait for at iteration $t$, and by $T_{i,t}$ the time interval between the update of the parameter vector $\vw_t$ at the PS and the reception of the $i$-th gradient $\vg_{i,t}$.

The general backup-workers scheme can be implemented in different ways with quite different performance.
When implementing the backup workers scheme, there are two general ways to synchronize  the PS and the workers: either the PS \emph{pushes} the updated parameter vector to workers or the workers \emph{pull} the most updated parameter vector from the PS. 
\paragraph{Pull \textrm{(Pl)}}
Whenever available to perform a new computation, a worker pulls the most updated parameter vector from the PS. Google's framework for distributed ML---TensorFlow 1.x~\cite{tensorflow}---implements Pl through a shared blocking FIFO queue of size $n$ where the PS enqueues $n$ copies of tokens indicating the corresponding iteration number. Whenever a worker becomes idle, it dequeues the token from the queue and retrieves the parameter vector directly from the PS.\footnote{We describe what appears to be an inefficient implementation. The parameter vector retrieved by the worker may correspond to a more recent iteration than what indicated in the token. Nevertheless, the corresponding gradient is still associated to the old iteration and then will be discarded at the PS. The worker may start then a computation that is already known to be useless!}

\paragraph{Push \& Interrupt \textrm{(PsI)}} 
After the PS updates the new parameter vector $\vw$, 
it pushes $\vw$ to all workers, 
which interrupt any ongoing computation to start computing a new gradient at $\vw$. 
%\emph{interrupts} the ongoing computation of the workers and sends (pushes) them $\vw$ to start new gradients' calculations. 
Interrupts can be implemented in different ways.
%e.g.~creating and destroying specific computation threads as in~\cite{teng18neurips}, or alternating computation steps and message listening in the same thread as in \cite{amiri2018computation}.
For example, in~\cite[Algo.~2]{teng18neurips}, the main thread at each worker creates a specific thread for each gradient computation and keeps listening for a new parameter vector. Once the worker receives the new one from PS, the computing thread is killed. 
However, the overhead of online creating/destroying threads is not negligible since it requires run-time memory allocation and de-allocation, which may even slow down the system~\cite{Ling:2000:AOT:346152.346320}.
In~\cite{amiri2018computation}, the same thread performs the computation but periodically checks for new parameter vectors from the PS. 
When the worker receives a new parameter vector, it stops its ongoing computation. 
The performance of this interrupt mechanism depends on how often  workers listen for messages from PS.%, which is not further discussed in the paper.

\paragraph{Push \& Wait \textrm{(PsW)}}
The PS pushes the new parameter vector to each worker as in PsI, but the worker 
completes its current computation 
before dequeueing the most recent parameter vector from a local queue. PsW
can be easily implemented using MPI non-blocking communication package~\cite{LiKAS18} or the FIFO queue provided in TensorFlow~\cite{LuoLZQ19}.

\bigskip
Our algorithm works with any of the variants listed above, with minor adaptations. 
We have implemented and tested it both with PsI and PsW in the PyTorch framework \cite{pytorch}. 
Results are similar, therefore, in what follows, we refer only to PsW.

To the best of our knowledge, there are two other proposals to dynamically adapt the number of backup workers~\cite{teng18neurips,dutta}. Both consider a PsI approach. In~\cite{teng18neurips} the PS uses a deep neural network to predict the time~$T_{k,t}$ needed to collect $k=1, 2, \dots n$ new  gradients. It then greedily chooses $k_t$ as the value that maximizes~$k/T_{k,t}$. This neural network for time series forecasting needs itself to be trained in advance for each cluster and each ML model to be learned. No result is provided in~\cite{teng18neurips} about the duration of this additional training phase or its sensitivity to changes in the cluster and/or ML models.
Our algorithm \DynamicBackUp{} also selects~$k_t$ to maximize a similar ratio, but 1) replaces the numerator by the expected decrease of the loss function, 2) uses a  simple estimator for~$T_{k,t}$, that does not require any preliminary training. 
%\DynamicBackUp{} has a much lower overhead: 1) no preliminary training is required, and 2) the computational. 
%We were not able to directly compare our approach with the one in~\cite{teng18neurips}, because the detailed description of the algorithm is in the supplementary material (currently not available). 
Moreover, results in~\cite{teng18neurips} do not show a clear advantage of the proposed mechanism in comparison to the static setting suggested in~\cite{Jianmin2016} (see~\cite[Fig.~4]{teng18neurips}). 
Our experiments in Sect.~\ref{s:experiments} confirm that indeed considering a gain proportional to~$k$ as in~\cite{teng18neurips} is too simplistic (and leads to worse results than \DynamicBackUp).
The recent paper~\cite{dutta} 
proposes \adasync{} that selects $k_t$ to minimize the average expected squared norm of the gradients over a time horizon. \adasync{} relies on an upper bound for the expected squared norm of the gradients and analytical formulas for $T_{k,t}$ for specific distributions of the computation times---they only develop the case for shifted exponential random variables. Finding the optimal $k_t$ would require to know or estimate at run-time some quantities like the Lipschitz constant or noise variance. \adasync{}  instead determines $k_t$ by solving an approximate quadratic equation that only depends on the current loss.
%equation as many values  some approximations are required considers a similar ratio upper-bound for the  that the numerator is replaced by the expected squared norm of the full gradients.  Meanwhile, a quadratic function is derived for choosing $k_t$, which requires a prior knowledge of the runtime distribution restrained to shifted exponential one to analytically have $T_{k,t}$. 
On the contrary, \DynamicBackUp{} estimates the different quantities online without prior information about the distribution of the computation times, and it is then able to adapt to changes in the cluster, e.g., due to   
%assumption on the propose estimators for the q
%Our algorithm \DynamicBackUp{} which provides an on-line estimator for $T_{k,t}$, does not require such a prior information and can adapt to any run-time distribution, even to the time-varying case due to the 
dynamic resource allocation (Sect.~\ref{subsec:robust}). 
When computation times are distributed according to a shifted exponential distribution, our experiments  show that \DynamicBackUp{} trains faster than \adasync{} when computation variability is small (Sect.~\ref{subsec:adasync}).

Our approach to estimate the loss decrease as a function of $k$ is inspired by the work~\cite{balles17uai} which evaluates the loss decrease as a function of the batch size. In fact, aggregating $k$ gradients, each computed on a mini-batch of $B$ samples, is almost equivalent to compute a single gradient on a mini-batch of $kB$ samples.
%(the gradient variance is larger due to possible resampling of the same dataset points at different workers).

While our algorithm adapts the number of backup workers $b$ given an available pool of $n$ workers, the authors of~\cite{wang19} proposes a reinforcement learning algorithm to adapt $n$ in order to minimize the training time under a budget constraint. This algorithm and \DynamicBackUp{} are then complementary: once selected $n$ with the approach in~\cite{wang19}, \DynamicBackUp{} can be applied to tune the number of backup workers.

\section{Dynamic backup workers}
\label{s:algorithm}

The rationale behind our algorithm \DynamicBackUp{} is to adaptively select $k_{t}$ in order to maximize $\frac{F(\vw_{t})-F(\vw_{t+1})}{T_{k,t}}$, i.e.,~to greedily  maximize the decrease of the empirical loss per time unit. We decide $k_t$ just after the update of $\vw_t$.\footnote{It is possible in principle to refine the choice of $k_t$ upon the arrival of the first gradients of $\vw_t$.}
In the following subsections, we detail how  both numerator and denominator can be estimated, % on the basis of previous iterations. 
and how they depend on $k$. 
The notation is listed in Table~\ref{tab:notation}.
%Our calculations highlights the effect of $k$ on both terms.

\begin{table}[H]
\centering
\begin{tabular}{l|l}
\hline
$t$ &  iteration number\\
$n$& number of workers\\
$\vw_t$ & parameter vector at iteration $t$\\
$F$ &(global) loss function to minimize  \\
$B$& batch size\\
$\eta$ & learning rate\\
$L$ & Lipschitz smoothness constant of $F$\\
%$\nabla F(\vw_t)$& full gradient of $F$ at $\vw_t$\\
$\vg_{i,t}$& $i^{th}$ stochastic gradient PS receives at iteration~$t$\\
$\V(\vg_{i,t})$& variance of $\vg_{i,t}$ \\
$k_t$& number of stochastic gradients  PS waits for at iteration~$t$\\
$\vg_t$& average gradient at iteration~$t$\\
$\mathcal G_{k,t}$& gain (expected loss decrease) if PS receives $k$ gradients\\ 
$T_{k,t}$ & time between $\vw_t$ update and $\vg_{k,t}$ reception at PS\\
$\mathsf{t}_{h,i,t}$& time between $\vw_t$ update and $\vg_{i,t}$ reception at PS\\
&  when PS has waited for $h$ gradients at iteration~$t-1$\\
$\mathcal{T}_{h,k}$& random variable from which $\mathsf{t}_{h,k,t}$ values are assumed to be sampled \\
$\sT_{h,k,t}$& set of  $\mathsf{t}_{h,k,t'}$ samples available up to iteration~$t$\\
\hline
\end{tabular}
\caption{Notation}
\label{tab:notation}
\end{table}

\subsection{Empirical Loss Decrease}\label{sec:loss_decrease}
We assume that the empirical loss function $F(\vw)$ is $L$-smooth, i.e.,~it exists a constant $L$ such that

\begin{align}
\label{e:lipschitz}
\lVert \nabla F(\vw') - \nabla F(\vw'') \rVert \le L \lVert \vw' - \vw'' \rVert, \forall \vw', \vw''.
\end{align}

Smoothness is a standard assumption in convergence results of gradient methods~(see for example~\cite{bubeck15,bottou18}). 
In our experiments we show \DynamicBackUp{} reduces the convergence time also when the loss is not a smooth function.
From ~\eqref{e:lipschitz} and~\eqref{e:update} it follows (see \cite[Sect.~4.1]{bottou18} for a proof):

\begin{align}
\label{e:lower_bound}
\Delta F_t & \triangleq F(\vw_t)- F(\vw_{t+1})  \ge \eta \nabla F(\vw_t)^\intercal \vg_t - \frac{L \eta^2}{2} \lVert \vg_t \rVert^2.
\end{align}

In order to select $k_t$, \DynamicBackUp{} uses this lower bound as a proxy for the loss decrease. 
We note, however, that $\vg_t$ depends on the value of $k_t$ (see \eqref{e:grad_bw}) and the random mini-batches  drawn at the workers. So at the moment to decide for $k_t$, $\vg_t$ is a random variable. We consider then the expected value (over the possible choices for the mini-batches) of the right-hand side of~\eqref{e:lower_bound}. We call it the \emph{gain} and denote by $\mathcal G_{k,t}$, i.e.,:

\begin{equation}
\label{e:gain_with_E}
    \mathcal G_{k,t}  \triangleq \E\left[\eta \nabla F(\vw_t)^\intercal \vg_t - \frac{L \eta^2}{2} \lVert \vg_t \rVert^2\right].
\end{equation}

Each stochastic gradient is an unbiased estimator of the full gradient, then $\E[\vg_t]= \nabla F(\vw_t)$. Moreover, for any random variable $X$, it holds $\E[X^2] = \E[X]^2 + \Var(X)$. Applying this relation to each of the component of the vector $\vg_t$, and then summing up, we obtain:

\begin{equation}
\label{e:second_moment}
\E[\lVert \vg_t \rVert^2]= \lVert \nabla F(\vw_t)\rVert^2 + \V(\vg_{i,t})/k,    
\end{equation}

where $\V(\vg_{i,t})$ denotes the sum of the variances of the different components of $\vg_{i,t}$, i.e.,~$\V(\vg_{i,t})\triangleq\sum_{l=1}^d \Var([\vg_{i,t}]_l)$. 
Notice that $\V(\vg_{i,t})$ does not depend on $i$, because each worker has access to the complete dataset. Then, combining~\eqref{e:gain_with_E} and~\eqref{e:second_moment}, $\mathcal G_{k,t}$ can be rewritten as 

\begin{align}
\mathcal G_{k,t} & =  \left(\eta-\frac{L \eta^2}{2} \right)\lVert \nabla F(\vw_t)\rVert^2 - \frac{L \eta^2}{2} \frac{\V(\vg_{i,t})}{k}. \label{e:exp_decrease}
\end{align}

\begin{figure*}[t]
%\vspace{.3in}
        \centering
        \begin{subfigure}[b]{0.31\textwidth}
            \centering
            \includegraphics[width=\textwidth]{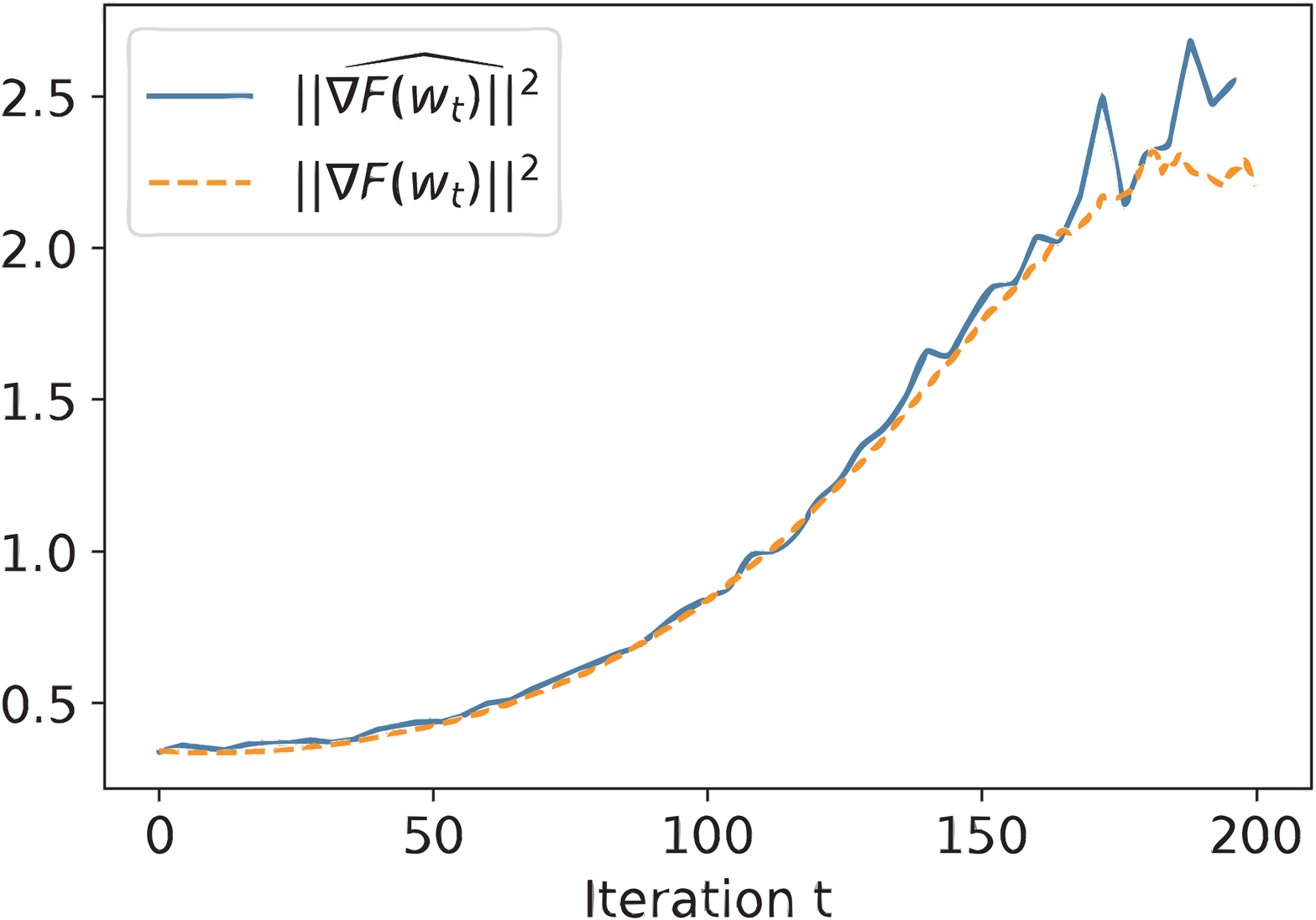}
                        \caption[]%
            {{\small Gradient norm}}    
            \label{f:gradientnorm}
        \end{subfigure}
        \hfill
        \begin{subfigure}[b]{0.31\textwidth}  
            \centering 
            \includegraphics[width=\textwidth]{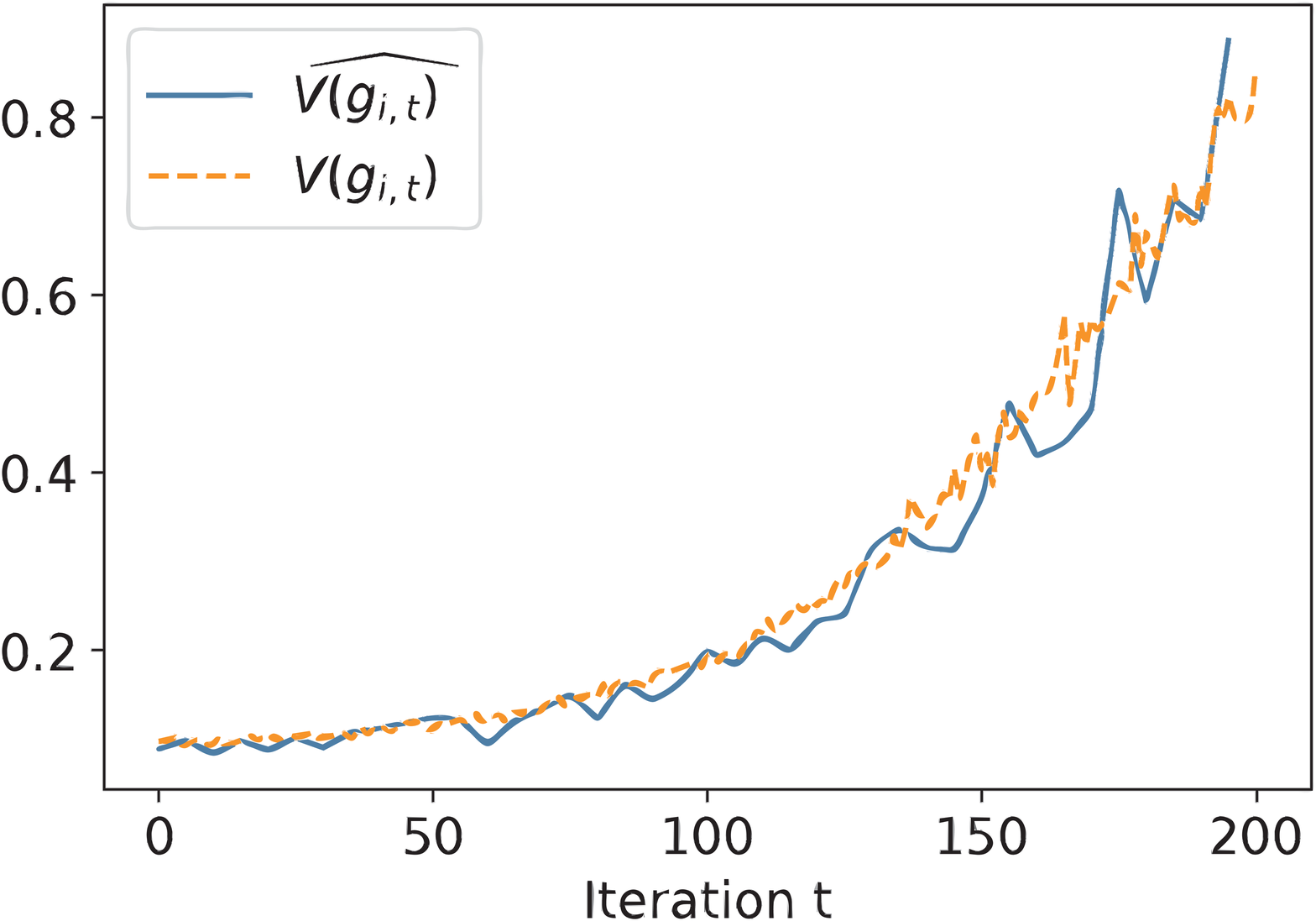}
                        \caption[]%
            {{\small Gradient variance}}    
            \label{f:variance}
        \end{subfigure}
        %\vskip\baselineskip
        \hfill
        \begin{subfigure}[b]{0.31\textwidth}   
            \centering 
            \includegraphics[width=\textwidth]{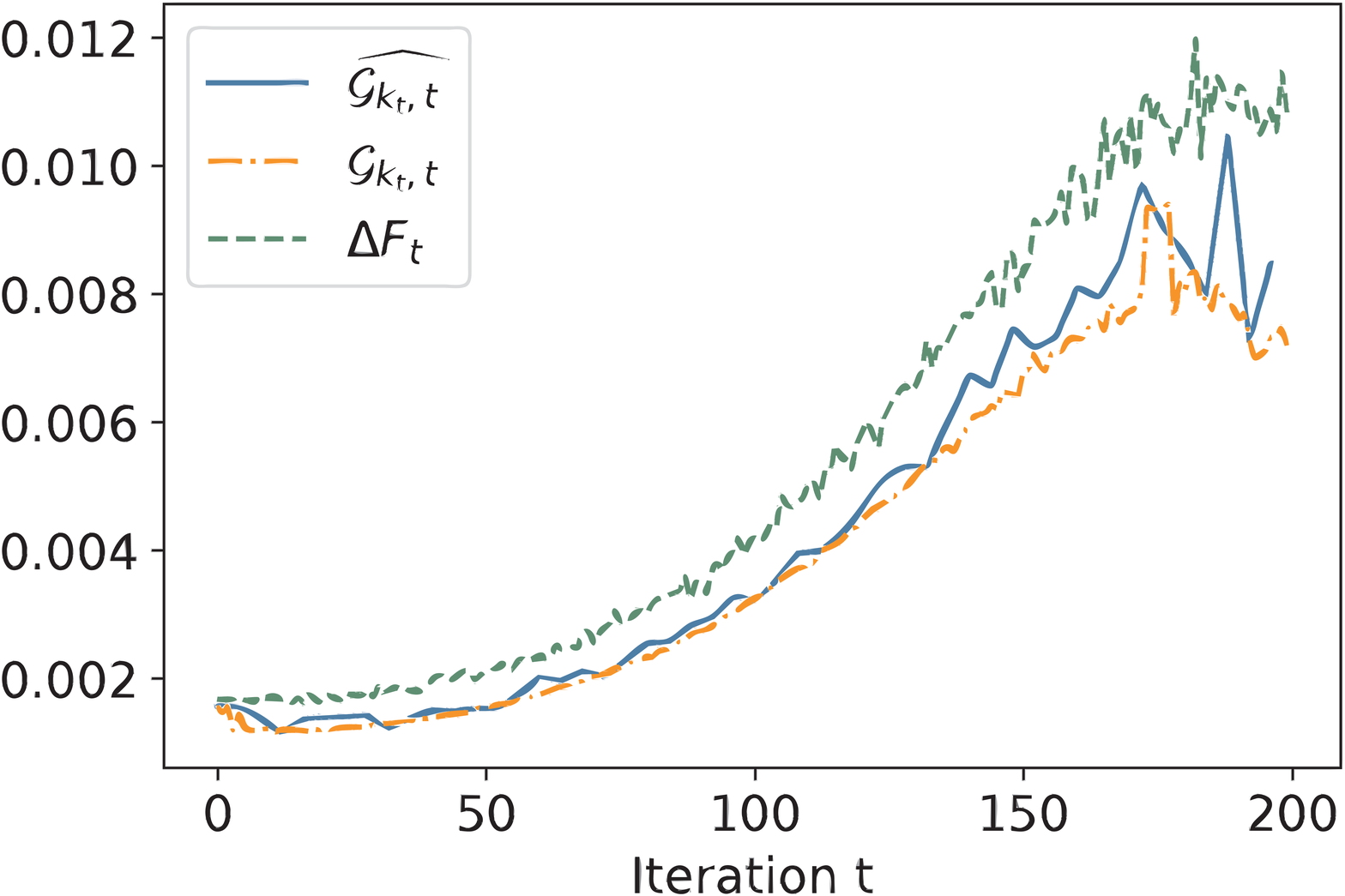}
            \caption[]%
            {{\small Loss decrease}}    
            \label{f:gain}
        \end{subfigure}
         \caption[ The average and standard deviation of critical parameters ]
        {\small Estimation of the loss decrease. MNIST, $n=16$ workers, batch size $B=500$, learning rate $\eta=0.01$, estimates computed over the last $D=5$ iterations.} 
        \label{Fig:Estimators}
    \end{figure*}

Equation~\eqref{e:exp_decrease} shows that the gain increases as $k$ increases. This corresponds to the fact that the more gradients are aggregated at the PS, the closer the stochastic gradient $-\vg_t$ is to its expected value $-\nabla F(\vw_t)$, i.e.,~to the steepest descent direction for the loss function. We also remark that the gain sensitivity to $k$ depends on the relative ratio of $\V(\vg_{i,t})$ and $\lVert \nabla F(\vw_t)\rVert^2$, that keeps  changing during the training (see for example Fig.~\ref{Fig:Estimators}). Correspondingly, we can expect that the optimal value of $k$ will vary during the training process, even when computation and communication times do not change in the cluster. Experiments in~Sect.~\ref{s:experiments} confirm this point.

Computing the exact value of  $\mathcal G_{k,t}$ would require the workers to process the whole dataset, leading to much longer iterations. 
%We need then to estimate 
%Notice that for machine learning problems, the calculation of empirical loss's gradient over all the data points ($\nabla F(\vw_t)$) takes a long time due to the size of the data set. Moreover, the variance of mini-batch gradient $\V(\vg_{i,t})$ is unknown and hard to compute. It is unpractical to get the exact value of $\mathcal G_{k,t}$. 
We want rather to evaluate $\mathcal G_{k,t}$  with limited overhead for the workers.
In what follows, we discuss how to estimate $\lVert\nabla F(\vw_t)\lVert^2$, $\V(\vg_{i,t})$, and $L$ to approximate $\mathcal{G}_{k,t}$ in~\eqref{e:exp_decrease}. We first provide estimators that use information available \emph{at the end} of iteration $t$, i.e.,~after $k_t$ has been selected and the $k_t$ fastest gradients have been received. Then, we build from these estimators new ones, that can be computed \emph{at the beginning} of the iteration $t$ and then can be used to select $k_t$. Given a quantity $\theta_t$ to be estimated at iteration $t$, we denote the first estimator as $\widehat{\theta_t}^+$ and the second one as $\widehat{\theta_t}$. 

We start by estimating $\V(\vg_{i,t})$ through the usual unbiased estimator for the variance:

\begin{align}
    \widehat{\V(\vg_{i,t})}^+ = \sum_{l=1}^d \frac{1}{k_t-1} \sum_{j=1}^{k_t} \left([\vg_{j,t} - \vg_t]_l \right)^2.
\label{e:sumvariancekt}
\end{align}

%The right hand side can exhibit quite high variability specially for small $k_t$. 
It is possible to have more precise estimates (even when $k_t=1$), if each worker can estimate $\V(\nabla f(x,\vw_t))$ from its mini-batch. As GPUs' low-level APIs do not provide access to such information, we do not further develop the corresponding formulas here.

Next, we study the estimator of $\lVert\nabla F(\vw_t)\lVert^2$. % supposing that PS receives $k_t$ gradients at iteration $t$. 
First, we can trivially use $\lVert \vg_t\rVert^2$ to estimate $\E[\lVert \vg_t\rVert^2]$, i.e.,~$\widehat{\E[\lVert \vg_t\rVert^2]}^+$ $= \lVert \vg_{t} \rVert^2$.
Since  $\lVert \nabla F(\vw_t)\rVert^2 = \E[\lVert \vg_t \rVert^2] - \V(\vg_{i,t})/k_t$ (from~\eqref{e:second_moment}), we can estimate $\lVert \nabla F(\vw_t)\rVert^2$ as follows

\begin{equation}
    \widehat{\lVert \nabla F(\vw_t)\rVert^2}^+\!\! = \max\!\left(\widehat{\E[\lVert \vg_t\rVert^2]}^+ -  \frac{\widehat{\V(\vg_{i,t})}^+}{k_t}, 0  \right),
\label{e:normsquarekt}
\end{equation}

where the $\max$ operation guarantees non-negativity of the estimate.

To estimate $L$, we need also to estimate $\mathcal G_{k_{t-1},t-1}$.  
In most of the existing implementations of distributed gradient methods for ML (including PyTorch's one), each worker~$i$ can send to the PS the local average loss computed on its mini-batch. % with tiny overhead. 
The PS can thus estimate the loss as

\[\widehat{F_t} = \frac{1}{k_t} \sum_{i=1}^{k_t} \frac{1}{B} \sum_{x \in \sB_i} h(x, \vw_t).\]

Thus, we have
\begin{equation*}
    \widehat{\mathcal G}_{k_{t-1},t-1}^+  = \widehat{F}_{t-1} - \widehat{F_t},
\end{equation*}
and substituting it to the left of \eqref{e:exp_decrease}, we get:
\begin{equation}
     \widehat{L_{t}}^+ = \frac{2 \left(\eta \widehat{\lVert \nabla F(\vw_{t-1})\rVert^2}^+ - \widehat{\mathcal G}_{k_{t-1},t-1}^+\right)}{\eta^2\left( \widehat{\lVert \nabla F(\vw_{t-1})\rVert^2}^+  + \widehat{\V(\vg_{i,t-1})}^+/k_{t-1} \right)} 
     \label{e:estimateL}
\end{equation}

Estimates in~\eqref{e:sumvariancekt}, \eqref{e:normsquarekt} and \eqref{e:estimateL} cannot be computed at the beginning of iteration $t$, but it is possible to compute them for earlier iterations, and use these past estimates to predict the future value. \DynamicBackUp{} simply averages the past $D$ estimates (or the first $t-1$ if $t\le D$), i.e.,

\begin{align}
\widehat{\V(\vg_{i,t})} & = \frac{1}{D} \sum_{v=1}^D \widehat{\V\left(\vg_{i,t-v}\right)}^+, \label{e:varianceoverD}\\
%\end{equation}
%and
%\begin{equation}
\widehat{\lVert \nabla F(\vw_t)\rVert^2} & = \frac{1}{D} \sum_{v=1}^D  \widehat{\lVert \nabla F(\vw_{t-v})\rVert^2}^+, \label{e:normsquareoverD} \\
 \widehat{L_{t}} & =  \frac{1}{D} \sum_{v=1}^{D} 
  \widehat{L_{t-v}}^+.
\label{e:LoverD}
%\end{equation}
\end{align}

Combining~\eqref{e:exp_decrease},~\eqref{e:varianceoverD},~\eqref{e:normsquareoverD} and~\eqref{e:LoverD}, the estimate of the gain is 
%denoted by $\widehat{\mathcal G_{k,t}}$, 

\begin{equation}
\label{e:gain}
\widehat{\mathcal G_{k,t}} = 
\left(\eta-\frac{ \widehat{L_{t}} \eta^2}{2} \right) \widehat{\lVert \nabla F(\vw_t)\rVert^2} - \frac{\widehat{L_{t}} \eta^2}{2} \frac{\widehat{\V(\vg_{i,t})}}{k}.
%\frac{\eta}{2}\left( \widehat{\lVert \nabla F(\vw_t)\rVert^2} - \frac{\widehat{\V(\vg_{i,t})}}{k} \right).
\end{equation}

     \begin{figure*}[t]
%\vspace{.3in}
        \centering
        \begin{subfigure}[b]{0.31\textwidth}
            \centering
            \includegraphics[width=\textwidth]{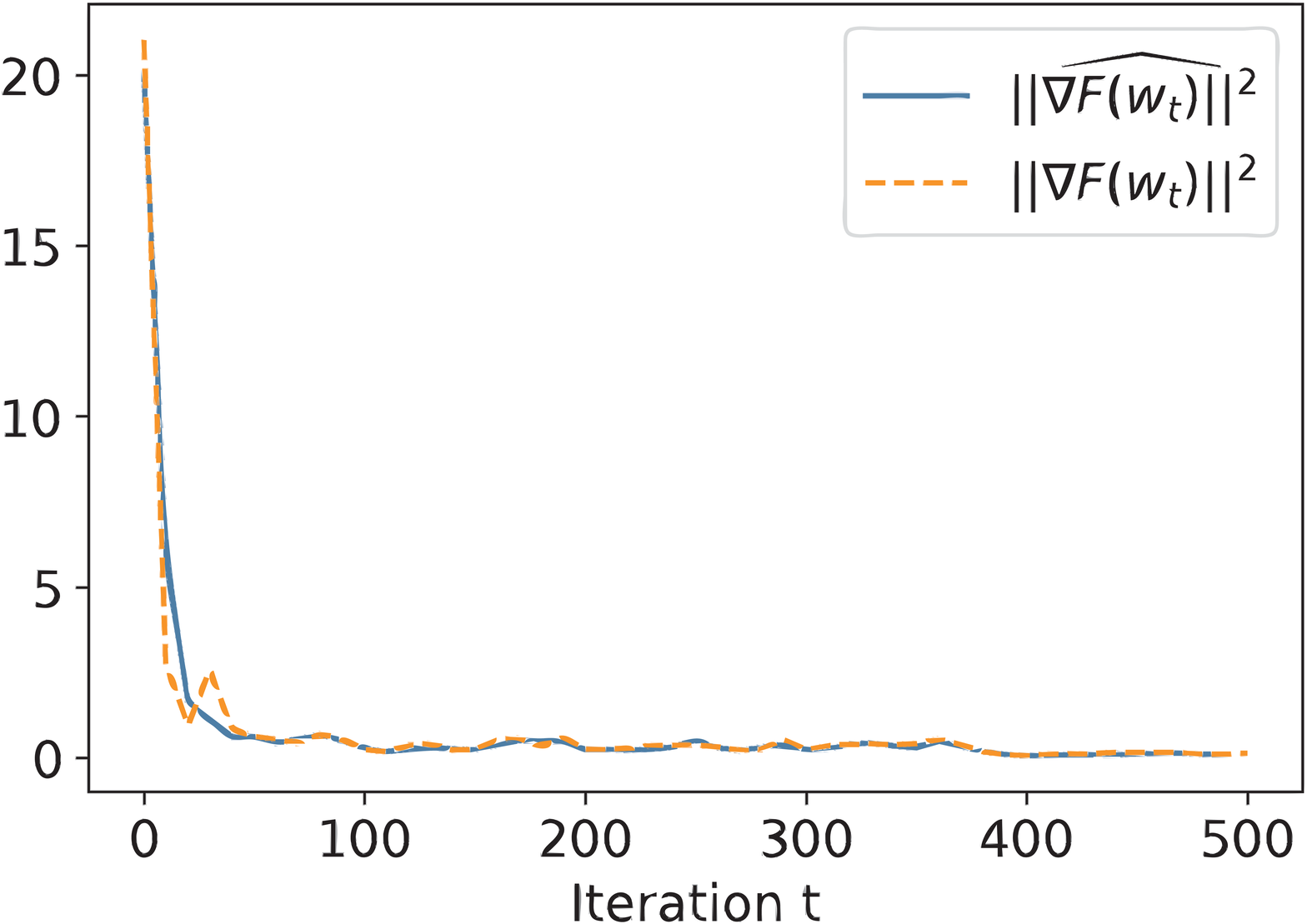}
                        \caption[]%
            {{\small Gradient norm}}    
            \label{f:gradientnorm_CIFAR}
        \end{subfigure}
        \hfill
        \begin{subfigure}[b]{0.31\textwidth}  
            \centering 
            \includegraphics[width=\textwidth]{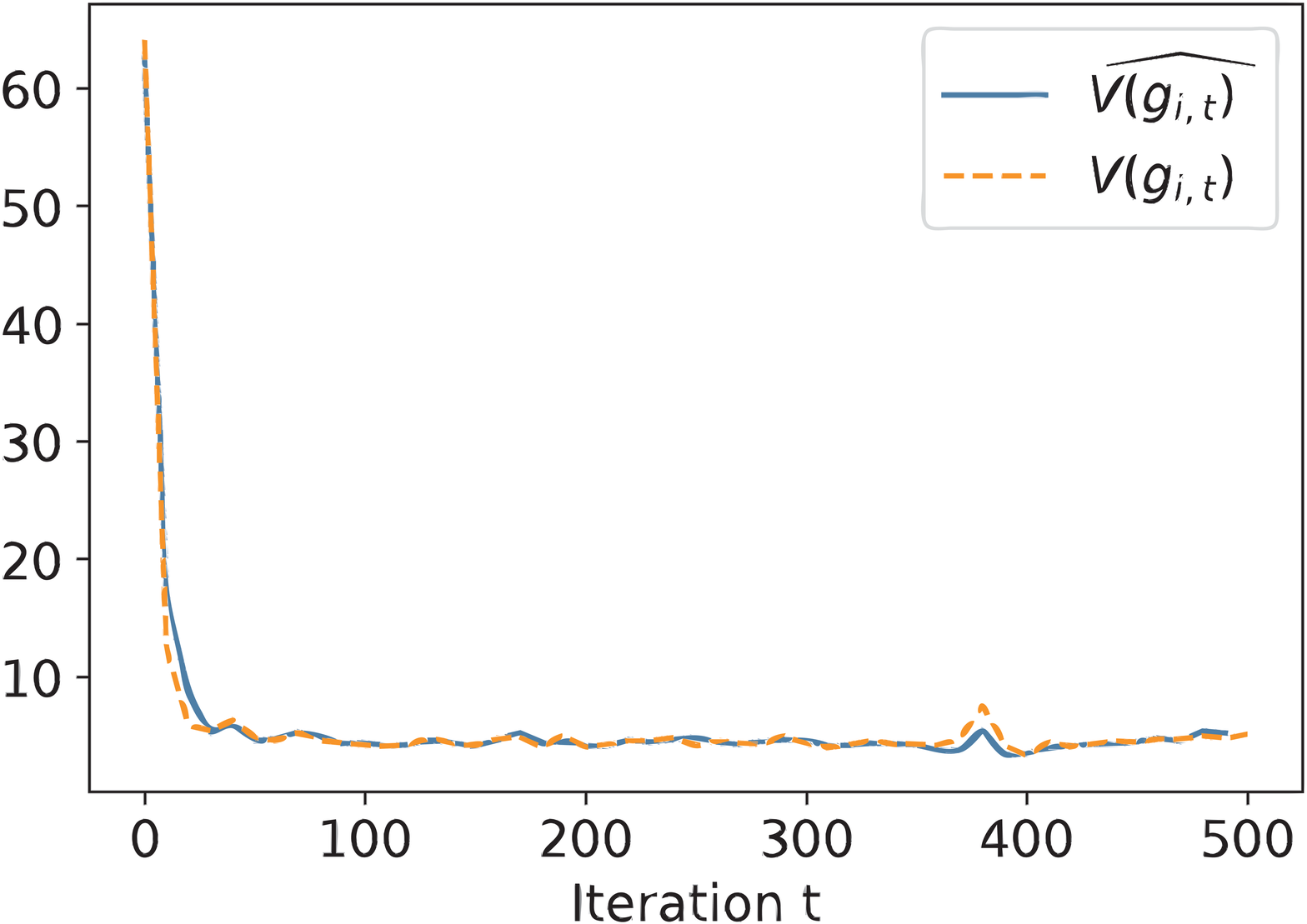}
                        \caption[]%
            {{\small Gradient variance}}    
            \label{f:variance_CIFAR}
        \end{subfigure}
        %\vskip\baselineskip
        \hfill
        \begin{subfigure}[b]{0.31\textwidth}   
            \centering 
            \includegraphics[width=\textwidth]{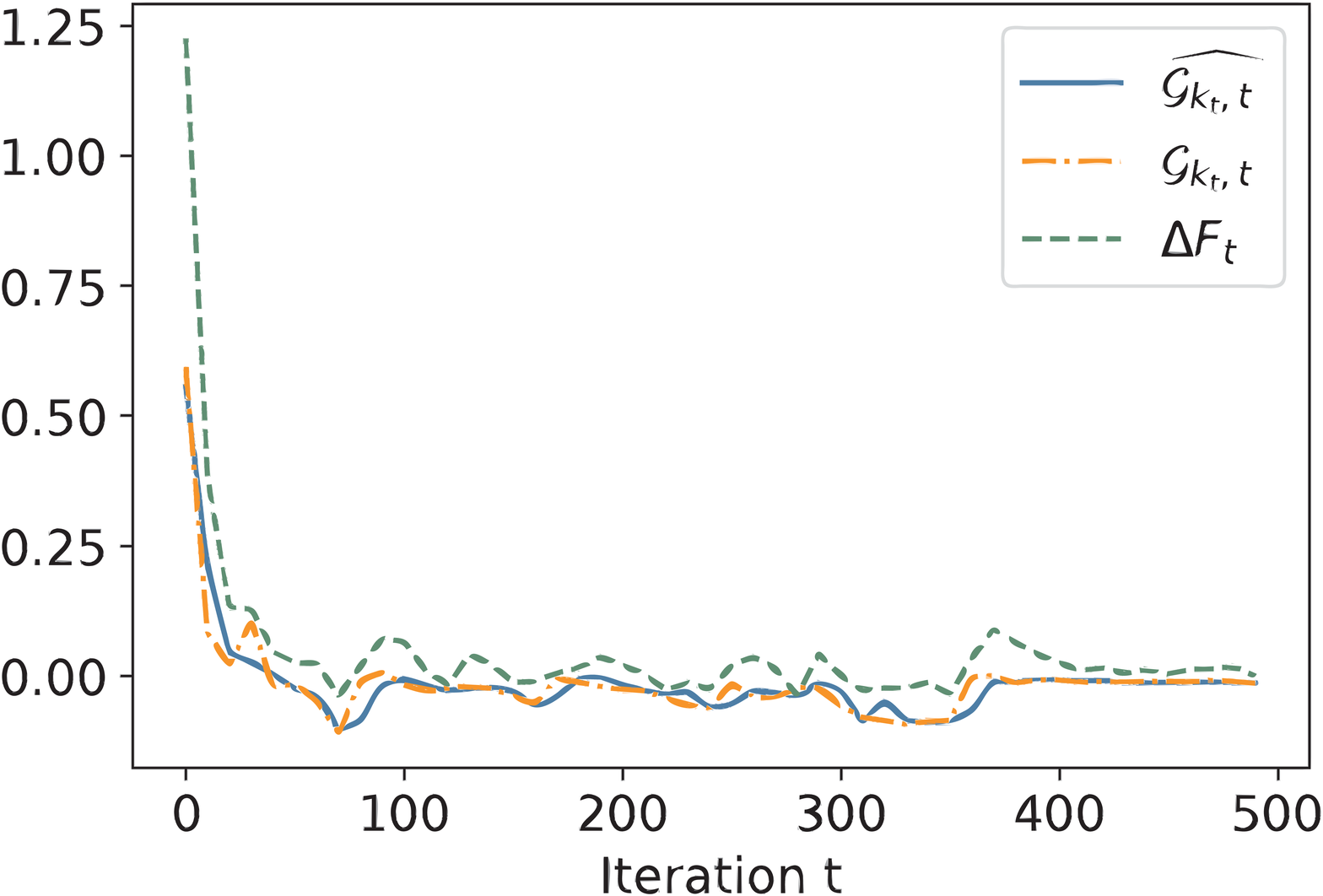}
            \caption[]%
            {{\small Loss decrease}}    
            \label{f:gain_CIFAR}
        \end{subfigure}
         \caption[ The average and standard deviation of critical parameters ]
        {\small Estimation of the loss decrease. CIFAR10, $n=16$ workers, batch size $B=256$, learning rate $\eta=0.05$, estimates computed over the last $D=5$ iterations.} 
        \label{Fig:Estimators_CIFAR}
    \end{figure*}

In Fig.~\ref{Fig:Estimators} and Fig.~\ref{Fig:Estimators_CIFAR}, we show our estimates during one training process on the MNIST and CIFAR10 dataset respectively (details in Sect.~\ref{s:experiments}), where our algorithm (described in Sect.~\ref{sec:select_k}) is applied to dynamically choose $k$. The solid lines are the estimates given by~\eqref{e:varianceoverD},~\eqref{e:normsquareoverD}, and~\eqref{e:gain}. The dashed lines present the exact values (we have instrumented our code to compute them). 
We can see from Figures~\ref{f:gradientnorm},~\ref{f:gradientnorm_CIFAR},~\ref{f:variance} and~\ref{f:variance_CIFAR} that the proposed estimates $\widehat{\lVert \nabla F(\vw_t)\rVert^2}$ and $\widehat{\V(\vg_{i,t})}$ are close to the true ones. Figures~\ref{f:gain} and~\ref{f:gain_CIFAR} compare the loss decrease $\Delta F_{t}$ (observed a posteriori) and  $\widehat{\mathcal{G}_{k_t,t}}$. As expected $\widehat{\mathcal{G}_{k_t,t}}$ is a lower bound for $\Delta F_{t}$, but the two quantities are almost proportional. This is promising, because, if the lower bound $\widehat{\mathcal{G}_{k,t}}/T_{k,t}$ and the function $\Delta F_{t}/T_{k,t}$ were exactly proportional, their maximizers would coincide. Then, working on the lower bound, as we do, would not be an approximation. Note that, for CIFAR10 dataset, the stochastic gradients are so noisy that the gradient variance is much larger than the gradient norm (as observed also in~\cite{DBLP:conf/aistats/NegliaXTC20}). Thus, the expected gain~\eqref{e:exp_decrease}, which is the lower bound for the loss decrease,  may become negative. 
In this case, \DynamicBackUp{} cautiously selects $k_t=n$ (see Sect.~\ref{sec:select_k}).

\subsection{Iteration Duration}\label{sec:time_estimation}
In this subsection, we discuss how to estimate the time $T_{k,t}$  the PS needs to receive $k$ gradients of $\vw_t$  after the update $\vw_t$ at iteration $t$. 
As in~\cite{Lee}, we call \emph{round trip time} the total (random) time an idle worker needs to 1)~retrieve the new parameter vector, 2)~compute the corresponding gradient, and 3)~send it back to the PS. %The PS measures round trip times to estimate $T_{k,t}$. 
Our estimators implicitly assume the cluster is stationary and homogeneous, in the sense that the distribution of round trip times does not change over time and from worker to worker. But in the experimental section, we show that they work also in dynamic and heterogeneous scenarios.

When the PS starts a new iteration $t$ ($t>0$), there are $k_{t-1}$ workers ready to compute the new gradient while the other $n-k_{t-1}$ workers are still computing stale gradients, i.e.,~relative to past parameter vectors~$\vw_{t-\tau}$ with $\tau>0$.
%In our design, $k_0=n$. 
$T_{k,t}$ depends not only on the value of $k$ but also on the value of $k_{t-1}$ and the $n-k_{t-1}$ residual round trip times (i.e.,~the remaining times for the $n-k_{t-1}$ busy workers to complete their tasks).
We assume that most of such dependence is captured by the number $k_{t-1}$. This would be correct if round trip times were exponential random variables due to their memoryless properties.
Let $\mathsf{t}_{h,i,t}$ denote the time the PS spends for receiving the $i$-th gradient of $\vw_t$, provided that it has waited $k_{t-1}=h$ gradients at iteration $t-1$. 
Under our assumptions, for given values of~$h$ and~$i$, the values~$\{\mathsf{t}_{h,i,t}\}$ can be seen as samples of the same random variable that we denote by~$\mathcal{T}_{h,i}$.
For estimating $T_{k,t}$, we consider $\widehat{T_{k,t}}= \widehat{\E[\mathcal{T}_{k,k}]}$.\footnote{
    It could seem more appropriate to consider $\widehat{T_{k,t}} = \widehat{\E[\mathcal{T}_{k_{t-1},k}]}$, but we want to select a value of $k$ that leads to good performance on the long term, i.e.,~if constantly used. For this reason, we use $\widehat{\E[\mathcal{T}_{k,k}]}$, that corresponds to select $k$ at each iteration.}

Consider $k_{t-1}=h$ and $k_t=k$. The PS can collect the samples $\mathsf{t}_{h,i,t}$ for $i \le k$ (it needs to wait $k$ gradients before moving to the next iteration), but also for $i>k$ because late workers still complete the ongoing calculations. 
In fact, late workers may terminate the computation and send their (by now stale) gradients to the PS, before they receive the new parameter vector. Even if a new parameter vector is available at the local queue (and then they know their gradient is not needed), in \DynamicBackUp{} workers still notify the completion to the PS, providing useful information to estimate $T_{k,t}$ with limited communication overhead.

A first naive approach to estimate $\E[\mathcal{T}_{k,k}]$ is to average the samples obtained over the past history. %This approach would lead to only use measurements from those iterations when the same value for $k$ has been selected  twice in a row. 
But, actually, there is much more information that can be exploited to improve estimations if we jointly estimate the complete set of values $\E[\mathcal{T}_{h,k}]$, for $h,k = 1, \dots n$.
In fact, the following pathwise relation holds for each $h$ and $i$: $\mathsf{t}_{h,i,t} \le \mathsf{t}_{h,i+1,t}$, because the index $i$ denotes the order of arrivals of the gradients. As a consequence, $\E[\mathcal{T}_{h,i}] \le \E[\mathcal{T}_{h,i+1}]$.  
Moreover, coupling arguments lead to conclude that $\E[\mathcal{T}_{h+1,i}] \le \E[\mathcal{T}_{h,i}]$ and $\E[\mathcal{T}_{i,i}] \le \E[\mathcal{T}_{i+1,i+1}]$. These two inequalities express the following intuitive facts: 1) if an iteration starts with more workers available to  compute, the PS will collect $i$ gradients faster (on average), {2) constantly waiting a smaller number of gradients leads to faster iterations.} As $\E[\mathcal{T}_{i,i}] \le \E[\mathcal{T}_{i+1,i+1}]$ may be less evident, we provide a proof in~\ref{a:proof}.
These inequalities allow us to couple the estimations of~$\E[\mathcal{T}_{h,k}]$, for $h,k = 1, \dots n$. Samples for a given pair $(h,k)$ can thus contribute not only to the estimation of~$\E[\mathcal{T}_{h,k}]$ but also to the estimations of other pairs. 
This is useful because the number of samples for $(h,k)$ is proportional to the number of times $k_t$ has been selected equal to $h$.
There can be many samples for a given pair and much less (even none) for another one.

Let $\sT_{h,k,t}$ be the set of samples available up to iteration $t$ for $(h,k)$, i.e.,~$\sT_{h,k,t}=\{\mathsf{t}_{h,k,t'}$, $\forall t'\le t \}$.
%We observe that the empirical average estimator is also the value that minimizes the mean squared error. 
We propose to estimate  $\{\E[\mathcal{T}_{h,k}], h, k = 1, \dots n \}$ by solving the following optimization problem:

\begin{align}
    \label{e:quadratic_opt}
    \underset{x_{h,k}}{\textrm{minimize}} \phantom{space} &   \sum_{h,k=1}^n \sum_{y \in \sT_{h,k,t}}(y -x_{h,k})^2\\
    \textrm{subject to}  \phantom{spac} &  x_{h,k} \le x_{h,k+1}, \;\;\;\textrm{   for }k=1, \dots n-1 \nonumber\\
                        &  x_{h+1,k} \le x_{h,k}, \;\;\; \textrm{ for }h=1, \dots n-1 \nonumber\\
                        &  x_{k,k} \le x_{k+1,k+1}, \textrm{ for }k=1, \dots n-1 \nonumber
\end{align}

Let $x_{h,k}^*$ be the solution of problem~\eqref{e:quadratic_opt}.
Then, $\widehat{\E[\mathcal{T}_{h,k}]} = x_{h,k}^*$, $\forall h,k = 1,\dots,n$ and we have  $\widehat{T_{k,t}}= x_{k,k}^*$.  We observe that, without the constraints, the optimal value $x_{h,k}^*$ at iteration $t$ is the empirical average of the corresponding set $\sT_{h,k,t}$. Hence, Problem~\eqref{e:quadratic_opt} is a natural way to extend the empirical average estimators, while accounting for the constraints.  
For our application, the quadratic optimization problem~\eqref{e:quadratic_opt} can be solved fast through solvers like CVX \cite{cvx,gb08} for the typical values of~$n$ ($10-1000$).

\begin{figure*}[t]
%\vspace{.3in}
        \centering
        \begin{subfigure}[b]{0.32\textwidth}   
            \centering 
            \includegraphics[width=\textwidth]{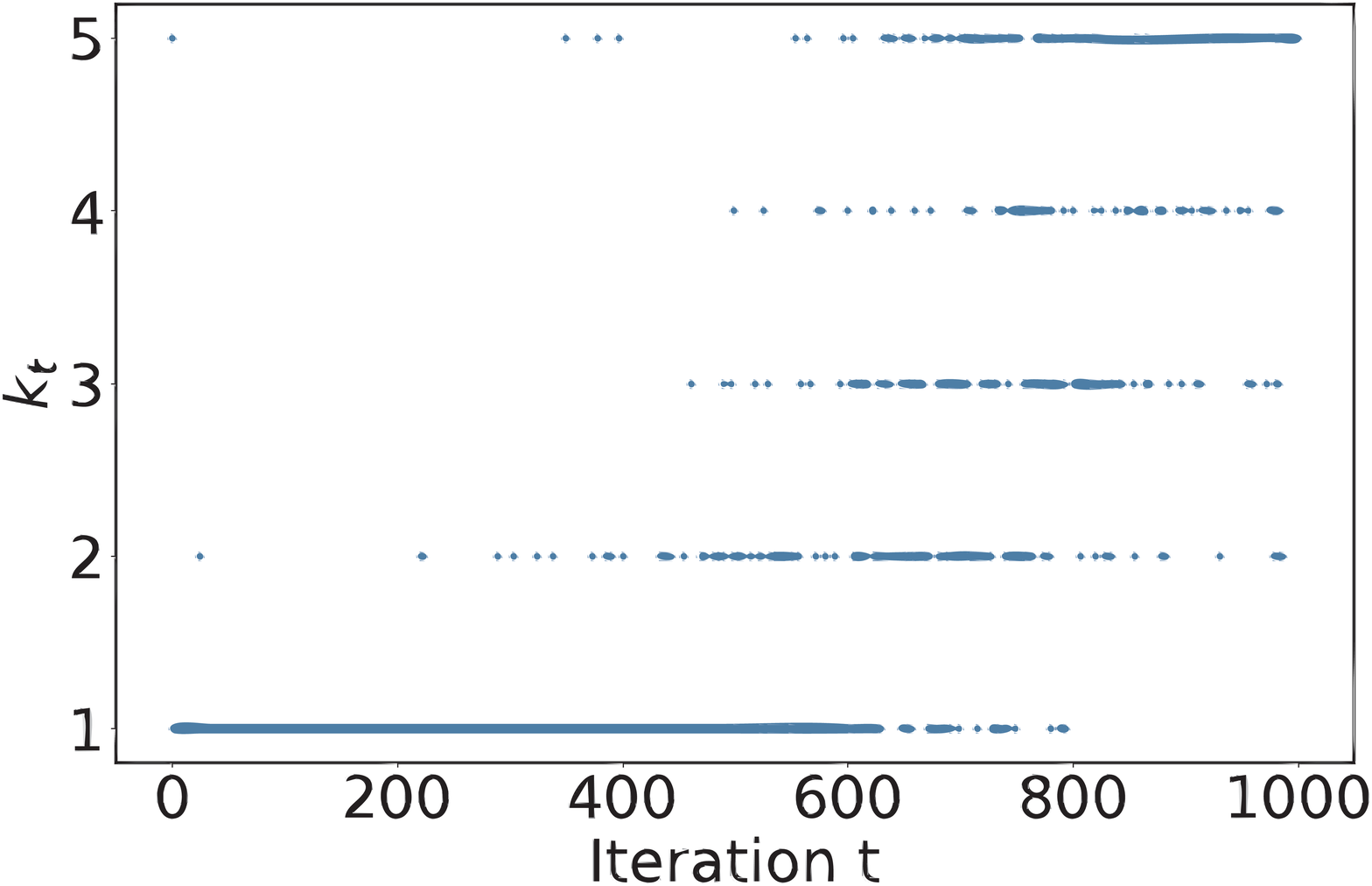}
                        \caption{\small Values of $k$ selected.}
                        %\label{Fig:TimeDistributionC}
        \end{subfigure}
        \hfill
        \begin{subfigure}[b]{0.32\textwidth}
            \centering
            \includegraphics[width=\textwidth]{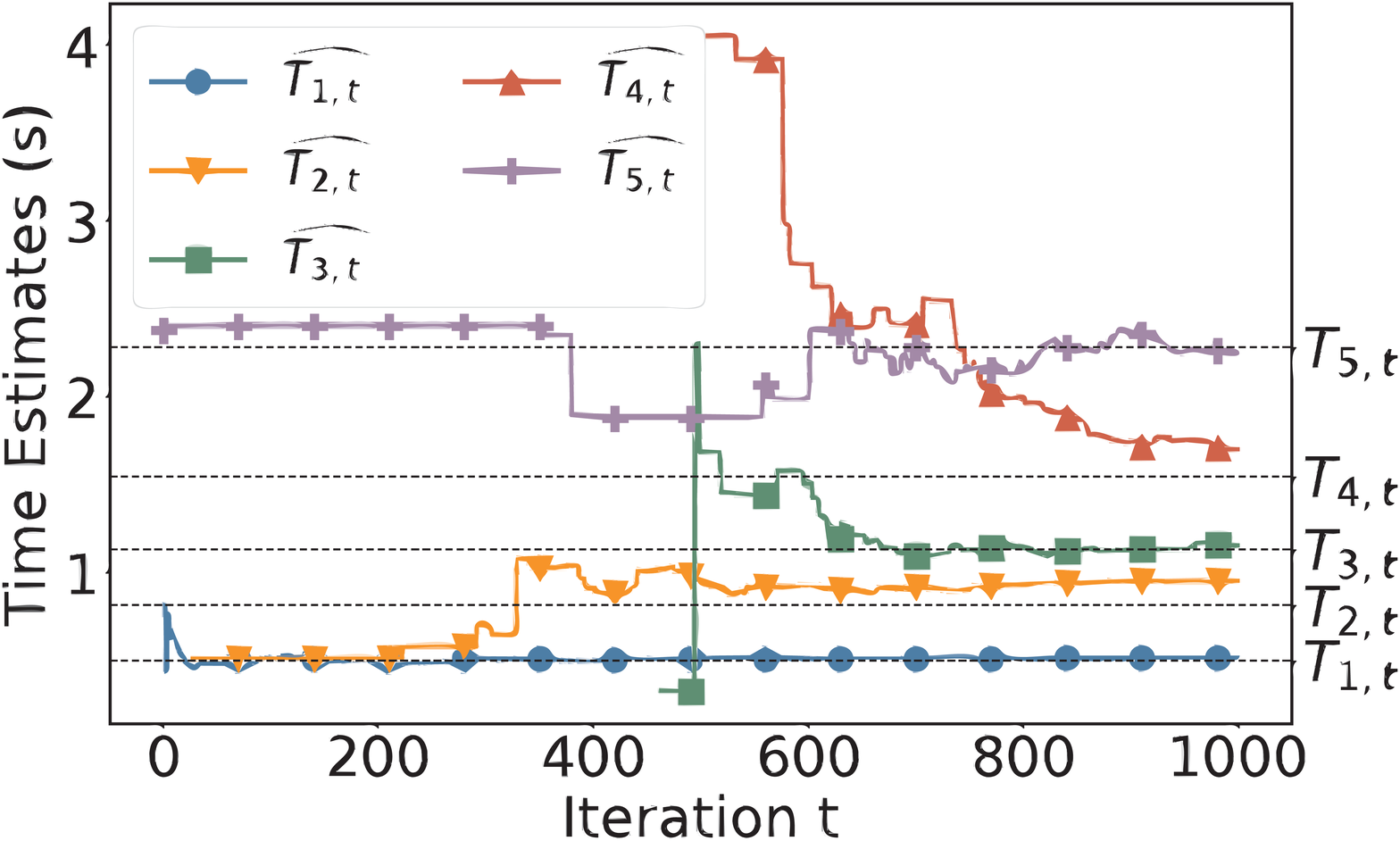}
                        \caption{\small Empirical average.}
        \end{subfigure}
        \hfill
        \begin{subfigure}[b]{0.32\textwidth}  
            \centering 
            \includegraphics[width=\textwidth]{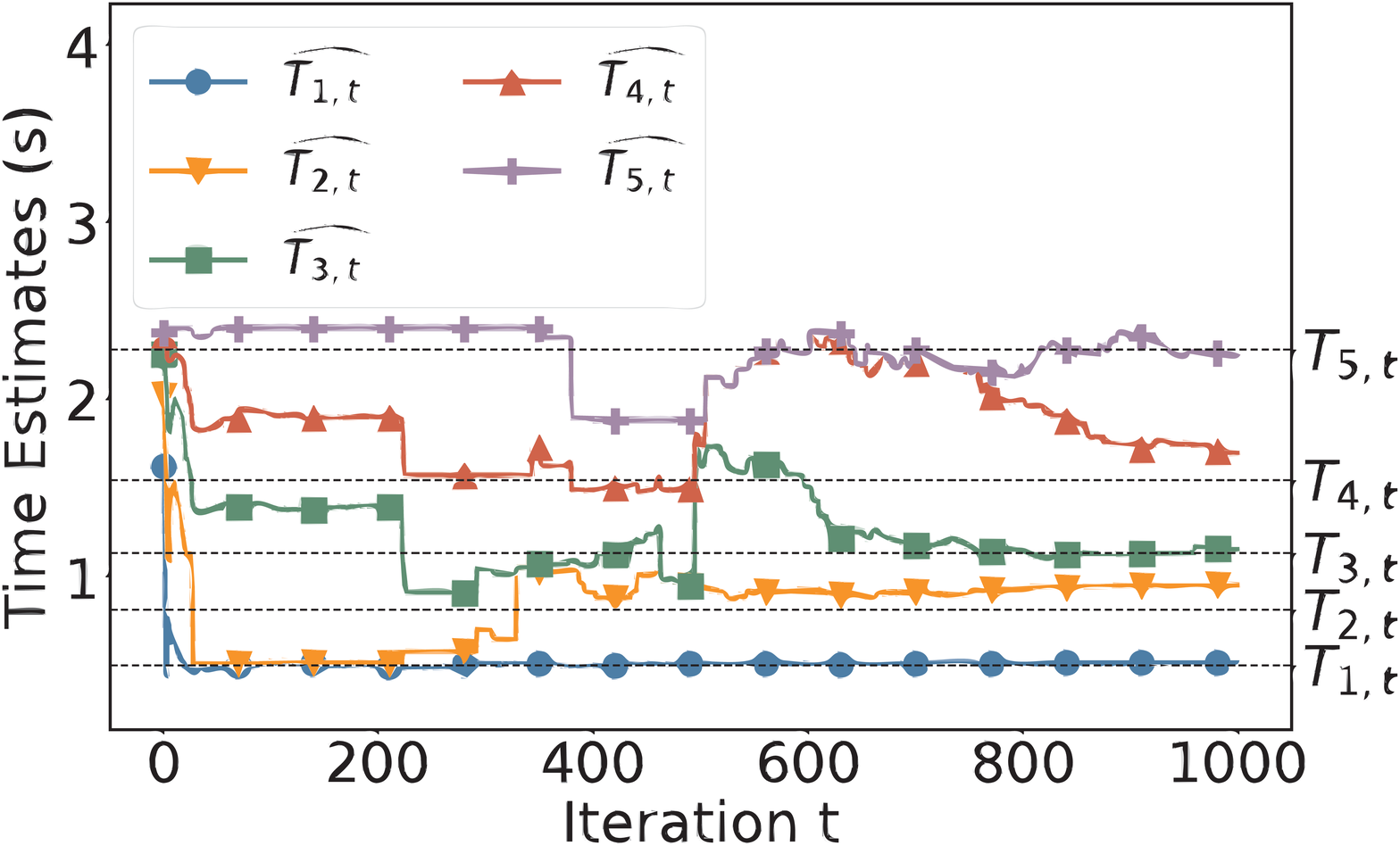}
                                    \caption{\small Constraint-aware estimator.}
        \end{subfigure}
        %\vskip\baselineskip
        \caption[ The average and standard deviation of critical parameters ]
        {\small Estimation of $T_{k,t}$. $n=5$ workers.} 
        \label{f:time_estimators}
    \end{figure*}

In Fig.~\ref{f:time_estimators}, we compare our estimator with the naive one (the empirical average).  We observe that the naive method 1) cannot provide estimates for a given value $h$ before it selects $k_t=h$, 2) leads often to  estimates that are in the wrong relative order.
By enforcing the inequality constraints, our estimator~\eqref{e:quadratic_opt} is able to obtain more precise estimates, in particular for the values $k=3$ and $k=4$ that are tested less frequently in this experiment. Experiments similar to those in Sect.~\ref{s:experiments} (but not shown in this paper) confirm that naive estimators lead to longer training time.

%\lipsum[1-2]

\subsection{Dynamic Choice of $k_t$}\label{sec:select_k}
\DynamicBackUp{} rationale is to select the parameter $k_t$ that maximizes the expected decrease of the loss function per time unit, i.e.,:

\begin{equation}
    k_t = \argmax_{1\le k \le n} \frac{\widehat{\mathcal G_{k,t}}}{\widehat{ T_{k,t}}}.
    \label{e:greedy}
\end{equation}

Note that \eqref{e:greedy} does not select values of $k$ for which $\widehat{\mathcal G_{k,t}}<0$, unless $\widehat{\mathcal G_{k,t}}<0$ for all values $k$, in which case 
%\eqref{e:greedy} selects 
$k_t=n$. 

This behaviour is correct. In fact,  $\widehat{\mathcal G_{k,t}}<0$ indicates the aggregate batch size $k B$ may be too low to guarantee that the stochastic gradient $\vg_t$ corresponds to a descent direction and then it is opportune to increase $k$ (if possible). Our approach then recovers some behaviour of dynamic sample size methods (see~\cite[Sect.~5.2]{bottou18}, \cite{de17aistats}). 
At the same time, $\mathcal G_{k,t}$ is a lower bound for the loss decrease $\E\left[\Delta F_t\right]$ (see~\eqref{e:lower_bound}). It may happen then that $\widehat{\mathcal G_{k,t}}<0$, even if  $\E\left[\Delta F_t\right]>0$. In this situation, \DynamicBackUp's choice of $k_t$ may not be optimal, as we observe in some settings in Sect.~\ref{subsec:adasync}, but still \DynamicBackUp{} errs on the side of caution to prevent the loss function from increasing. % to guarantee that the loss function decreases.

In addition, \DynamicBackUp{} exploits the local average loss $\widehat{F_t}$
to avoid decreasing $k_t$ from one iteration to the other, when the loss appears to be increasing (and then we need more accurate gradient estimates, rather than noisier ones). We modify~\eqref{e:greedy} to

\begin{equation}
    k_t = \max\!\left(\!\argmax_{1\le k \le n} \frac{\widehat{\mathcal G_{k,t}}}{\widehat{T_{k,t}}}, \; (k_{t-1}+1) \cdot \mathbbm{1}_{ \{\hat{F}_{t-1}>\beta \hat{F}_{t-2}\}  \wedge  \{k_{t-1}<n\} } \right),
    \label{e:greedy2}
\end{equation}

where $\beta \ge 1$ (we select $\beta=1.01$ in our experiments) and $\mathbbm{1}_A$ denotes the indicator function (equal to $1$ iff $A$ is true). If the loss has become $\beta$ times larger since the previous iteration, then \eqref{e:greedy2} forces $k_t \ge k_{t-1}+1$.

\section{Experiments}
\label{s:experiments}
We have implemented \DynamicBackUp{} in PyTorch~\cite{pytorch}, using the MPI backend for distributed communications. 
The experiments have been run on a real CPU/GPU cluster platform, with different GPUs available (e.g.,~GeForce GTX 1080 Ti, GeForce GTX Titan X, and Nvidia Tesla V100).
In order to have a fine control over the round trip times, our code can generate computation and communication times according to different distributions (uniform, exponential, Pareto, etc.) or read them from a trace provided as input file. The system operates at the maximum speed guaranteed by the underlying cluster, but it maintains a virtual clock to keep track of when events would have happened. Note that the virtual time is not a simple relabeling of the time axis: for example virtual time instants at which gradients are received by the PS determine which of them are actually used to update the parameter vector. So the virtual time has an effect on the optimization dynamics. Our code is available online~\cite{github_dbw}.

In what follows, we show that the number of backup workers should vary, not only with the round trip time distribution, but also with the hyper-parameters of the optimization algorithm like the batch size $B$. Moreover, the optimal setting depends as well on the stage of the training process, and then changes over time, even  when the cluster is stationary (round trip times do not change during the training period).

In all experiments, \DynamicBackUp{} achieves nearly optimal performance in terms of convergence time, and sometimes it even outperforms the optimal static setting, that is found through an exhaustive offline search over all values $k\in \{1,\dots,n\}$. %The algorithm is then quite robust. 
We also compare \DynamicBackUp{} with a variant where the gain  $\mathcal G_{k,t}$ is not estimated as in~\eqref{e:gain}, but it equals the number of aggregated gradients $k$, as proposed in~\cite{teng18neurips}. We call this variant blind \DynamicBackUp{} (\bdbw), because it is oblivious to the current state of the training. We find that this approach is too simplistic: ignoring the current stage of the optimization problem leads to worse performance  than~\DynamicBackUp.

We evaluated \DynamicBackUp{}, \bdbw{}, and different static settings for $k$ on two classification problems 1)~MNIST~\cite{mnist}, a dataset with $70000$ $28\times28$ images portraying handwritten digits from 0 to 9 and 2) CIFAR10~\cite{krizhevsky2009learning}, a dataset with $60000$ $32\times32$ colour images in 10 classes.\footnote{Both dataset include $10000$ test images.}
We trained a neural network with two convolutional layers with 5$\times$5 filters and two fully connected layers for MNIST and we trained a ResNet18~\cite{he2016deep} network for CIFAR10. The loss function was the cross-entropy one.
For MNIST, every worker had access to the entire dataset. For CIFAR10, the data set was split uniformly at random  among workers.

The learning rate is probably the most critical hyper-parameter in ML optimization problems. Ideally, it should be set to that largest value that still guarantees convergence. 
It is important to note that different static settings for the number of backup workers require different values for the learning rate. In fact, the smaller is $k$, the noisier is the aggregate gradient $\vg_t$, so that the smaller should be the learning rate. The rule of thumb proposed in the seminal paper~\cite{Jianmin2016} is to set the learning rate proportional to $k$, i.e.,~$\eta(k)\propto k$. This corresponds to the standard recommendation to have the learning rate proportional to the (aggregate) batch size~\cite{goyal17,smith18iclr}. 
In static settings,  aggregating $k$ gradients is equivalent to use a batch size equal to $kB$, so that the learning rate should scale accordingly. 
An alternative approach is to tune the learning rate independently for each static value of $k$ according to the empirical rule in~\cite{smith17}, that requires to run a number of experiments and determine the inflection points of a specific curve. 
This rule leads as well to learning rates increasing with $k$.
We call the two settings respectively the \emph{proportional} and the \emph{knee} rule. The maximum learning rate for the proportional rule is set equal to the value determined for $k_t=n$ by the knee rule.
%When using the proportional rule for static settings, the maximum learning rate $\eta(n)$ is set to be the value determined for $k_t=n$. 
The same value is also used as learning rate for  \DynamicBackUp{} and \bdbw{}, independently from the specific value they select for $k_t$. In fact, \DynamicBackUp{} and \bdbw{} can safely operate with a large learning rate because they dynamically increase $k_t$ up to $n$, when they detect that the loss is increasing. 
    
\begin{figure*}[t]
    \centering
    \begin{subfigure}[b]{0.48\textwidth}
        \centering
        \includegraphics[width=\textwidth]{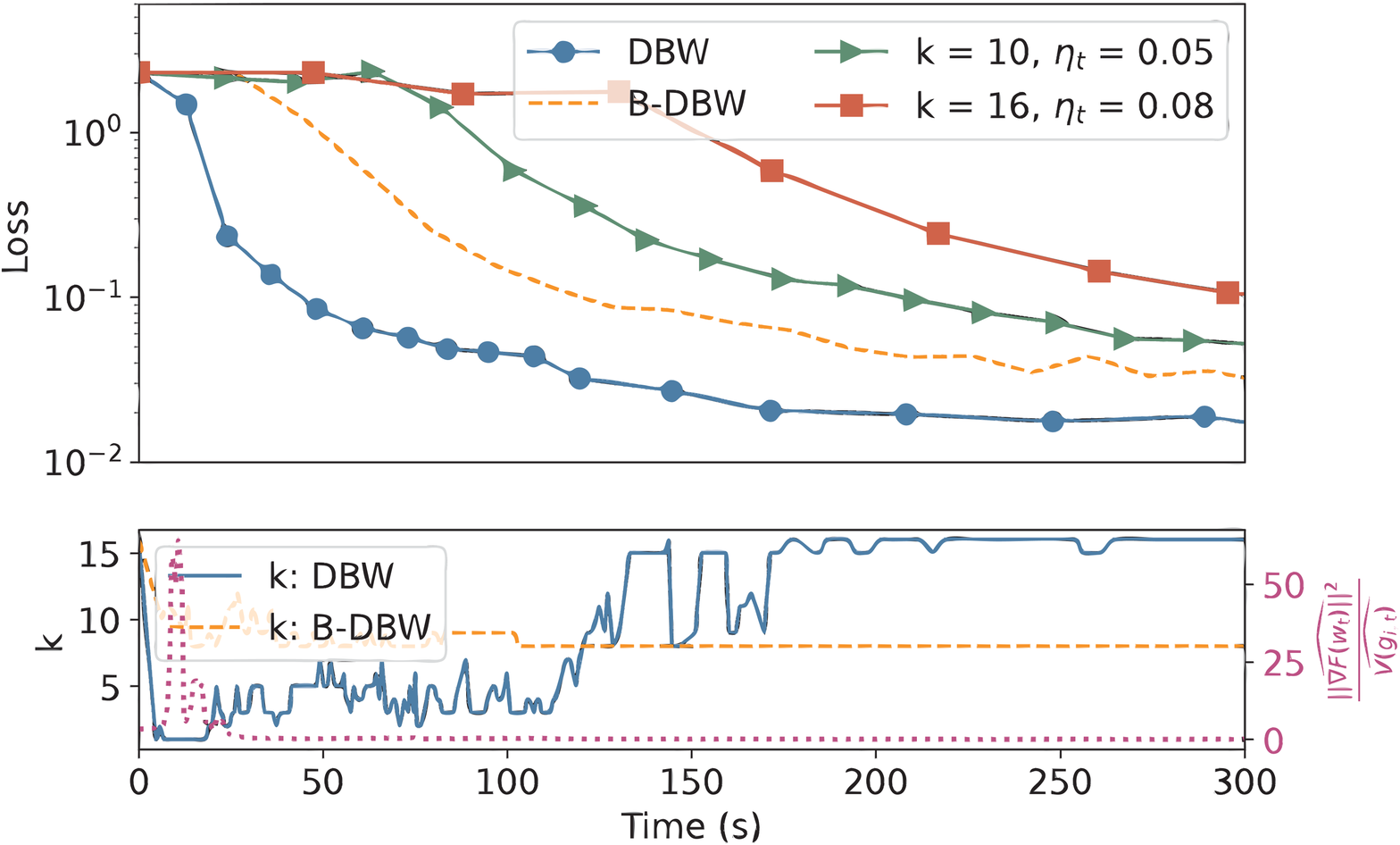}
        \caption{\small  Loss versus Time}
        \label{fig:one_run_loss}
    \end{subfigure}
    \begin{subfigure}[b]{0.48\textwidth}
        \centering
        \includegraphics[width=\textwidth]{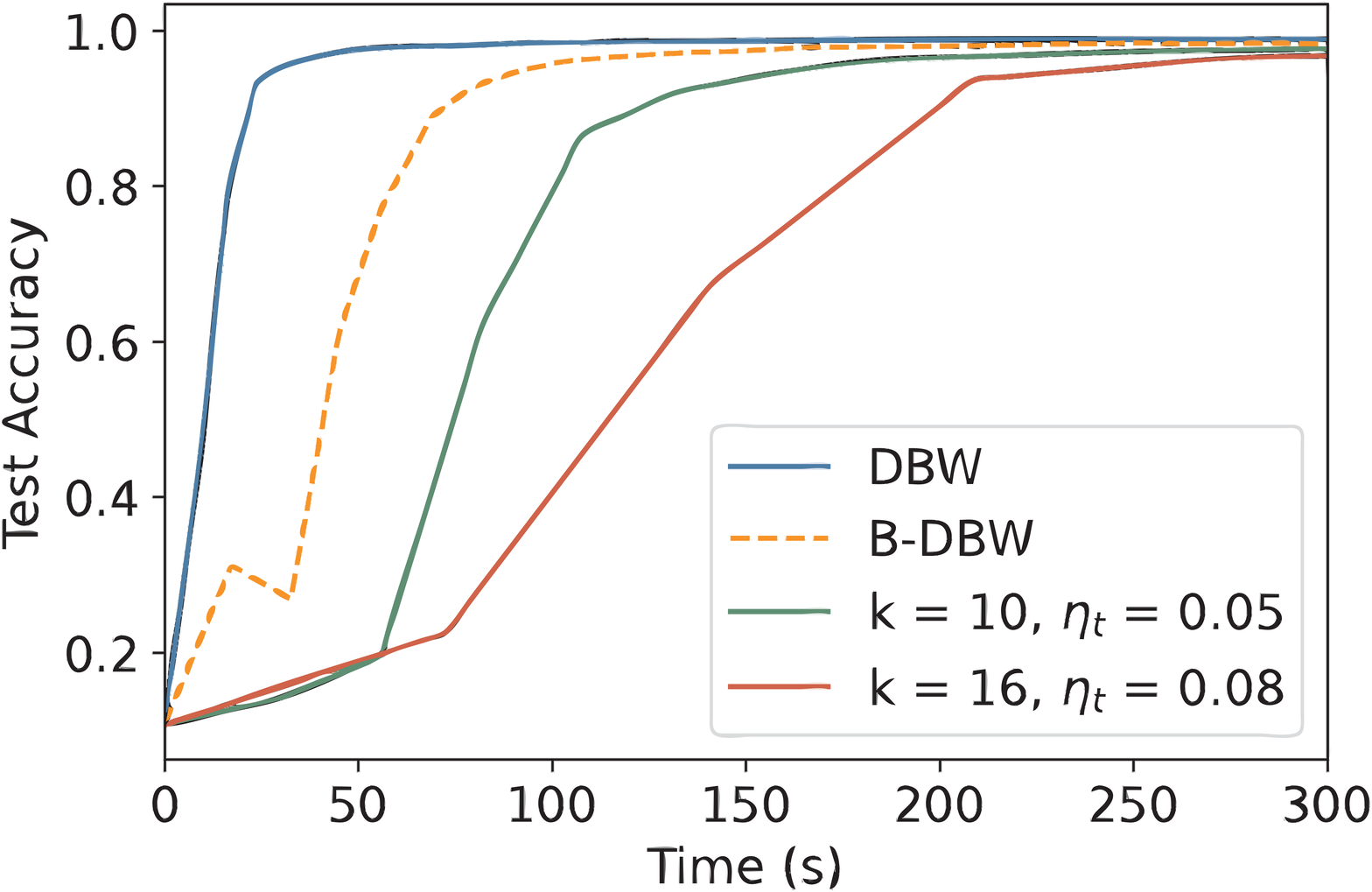}
         \caption{\small Accuracy versus Time}
         \label{fig:one_run_acc}
    \end{subfigure}
    \caption{Training on MNIST, batch size $B=500$, $n=16$ workers, estimates computed over the last $D=5$ iterations, proportional rule with $\eta(k)=0.005k$, round trip times follow shifted exponential distribution $0.3+0.7\textrm{Exp}(1)$.}
    \label{fig:one_run}
\end{figure*}
    
\begin{figure*}[t]
    \centering
    \begin{subfigure}[b]{0.48\textwidth}
        \centering
        \includegraphics[width=\textwidth]{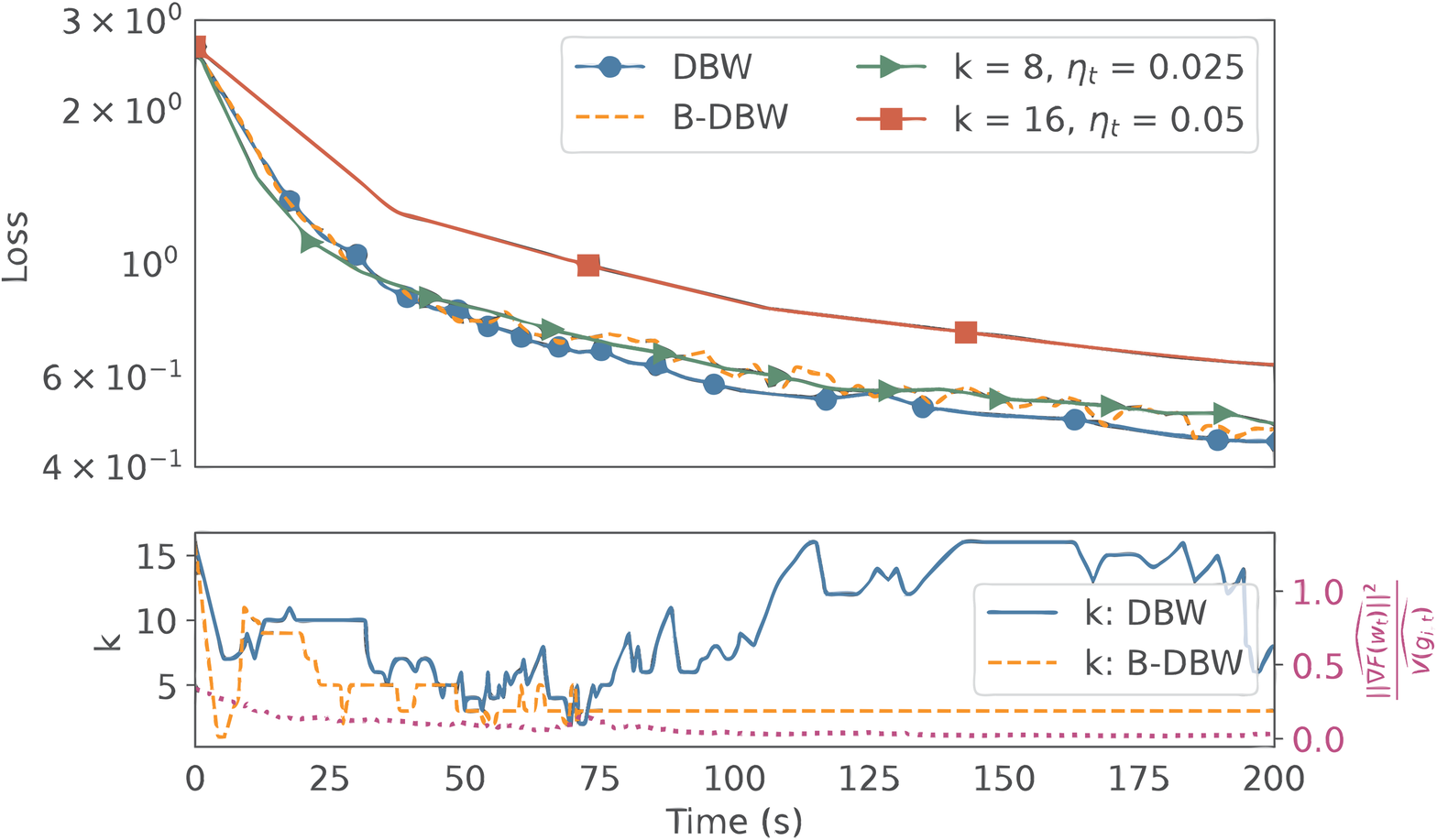}
        \caption{\small  Loss versus Time}
         \label{fig:one_run_cifar_loss}
    \end{subfigure}
    \begin{subfigure}[b]{0.48\textwidth}
        \centering
        \includegraphics[width=\textwidth]{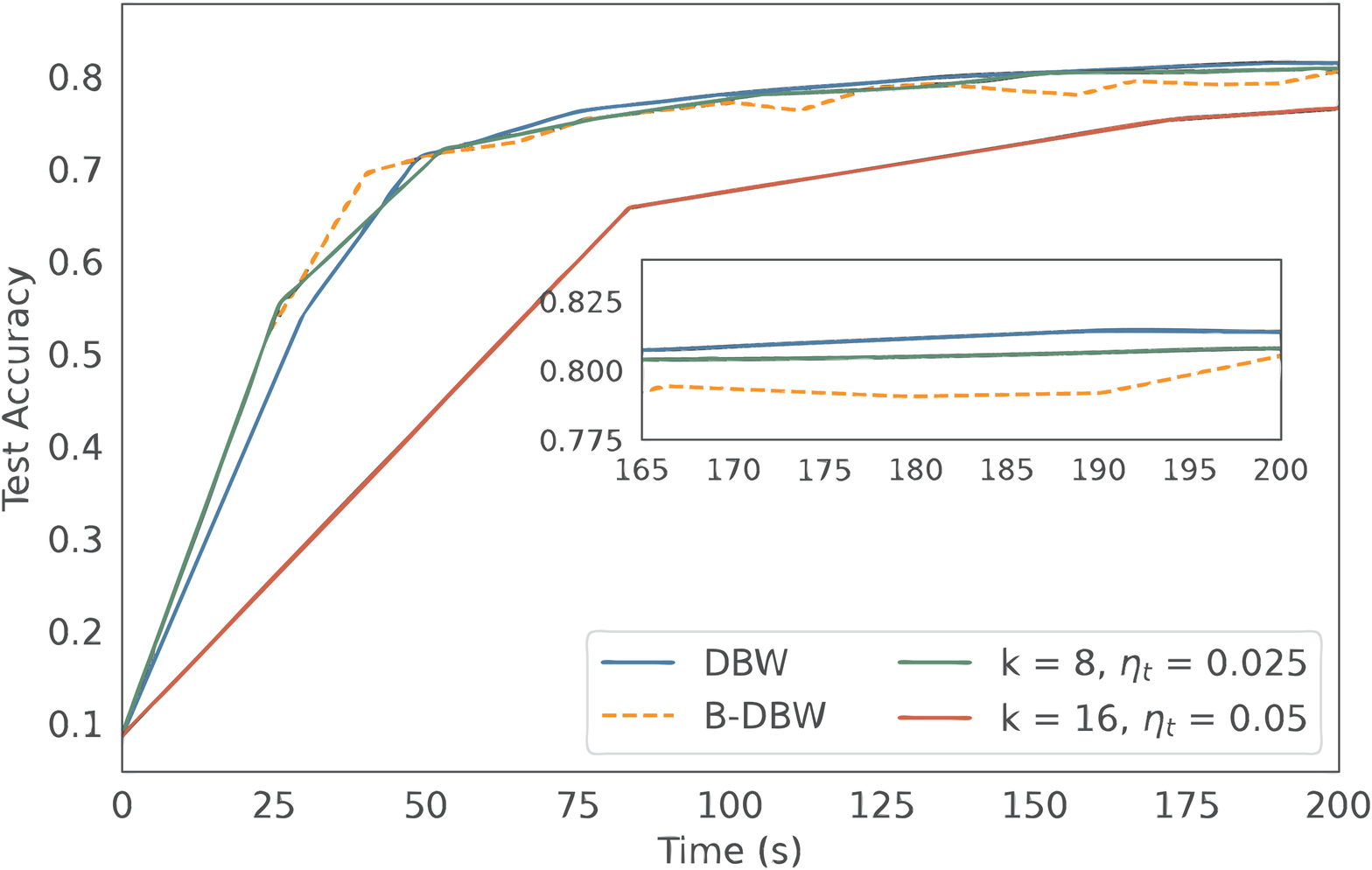}
         \caption{\small  Accuracy versus Time}
          \label{fig:one_run_cifar_acc}
    \end{subfigure}
    
        \begin{subfigure}[b]{0.48\textwidth}
        \centering
        \includegraphics[width=\textwidth]{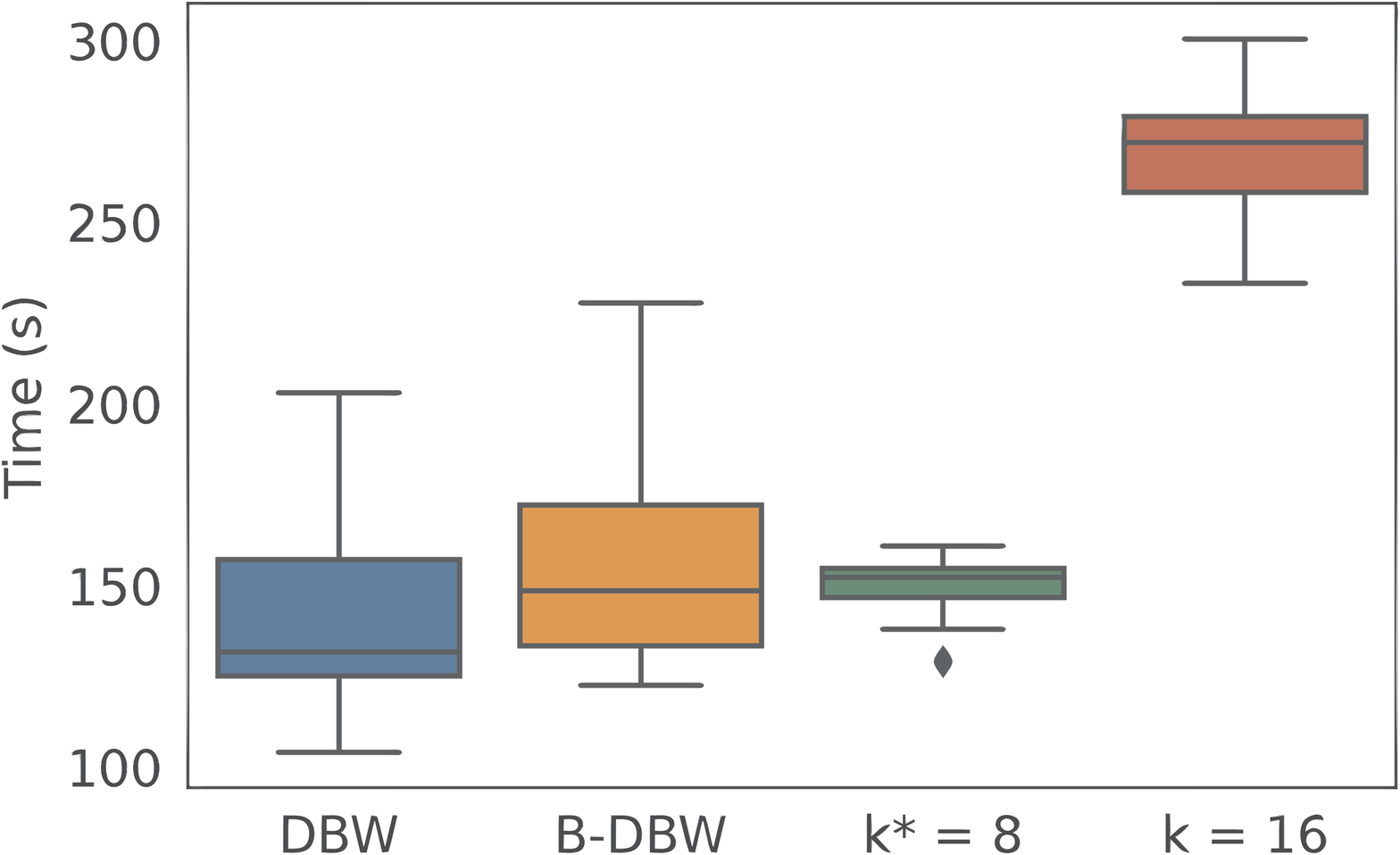}
        \caption{\small Time to reach $80\%$ test accuracy}
                 \label{fig:one_run_cifar_time_acc}
    \end{subfigure}
    \begin{subfigure}[b]{0.48\textwidth}
        \centering
        \includegraphics[width=\textwidth]{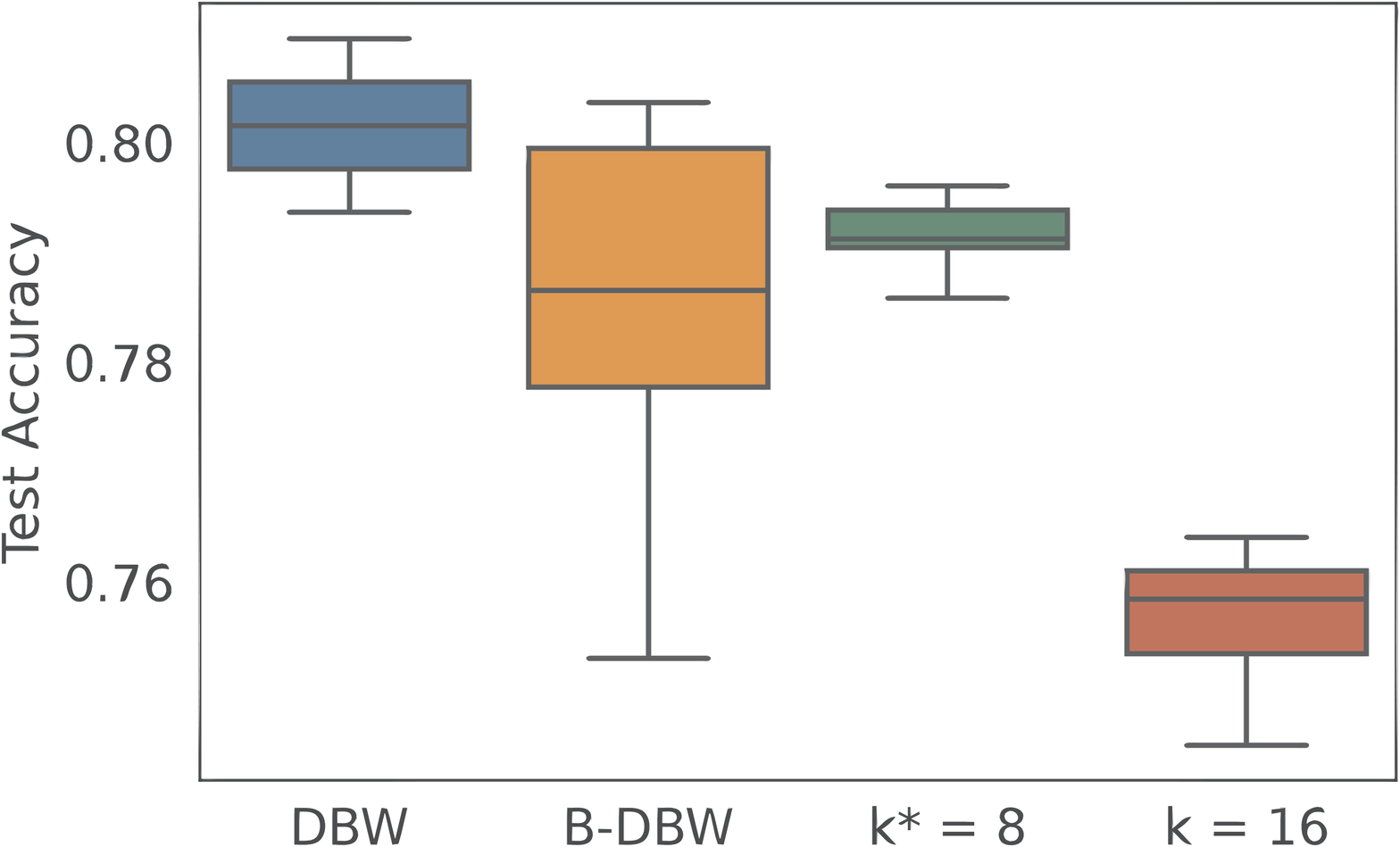}
         \caption{\small Test accuracy after 200 seconds}
         \label{fig:one_run_cifar_acc_time}
    \end{subfigure}
    \caption{Training on CIFAR10, batch size $B=256$, $n=16$ workers, estimates computed over the last $D=5$ iterations, proportional rule with $\eta(k)=\frac{0.05k}{16}$, round trip times follow exponential distribution $\textrm{Exp}(1)$. Box plots are bases on 20 independent runs.}
    \label{fig:one_run_cifar}
\end{figure*}

%\lipsum[1-2]
    \begin{figure*}[t]
%\vspace{.3in}
        \centering
        \begin{subfigure}[b]{0.32\textwidth}
            \centering
            \includegraphics[width=\textwidth]{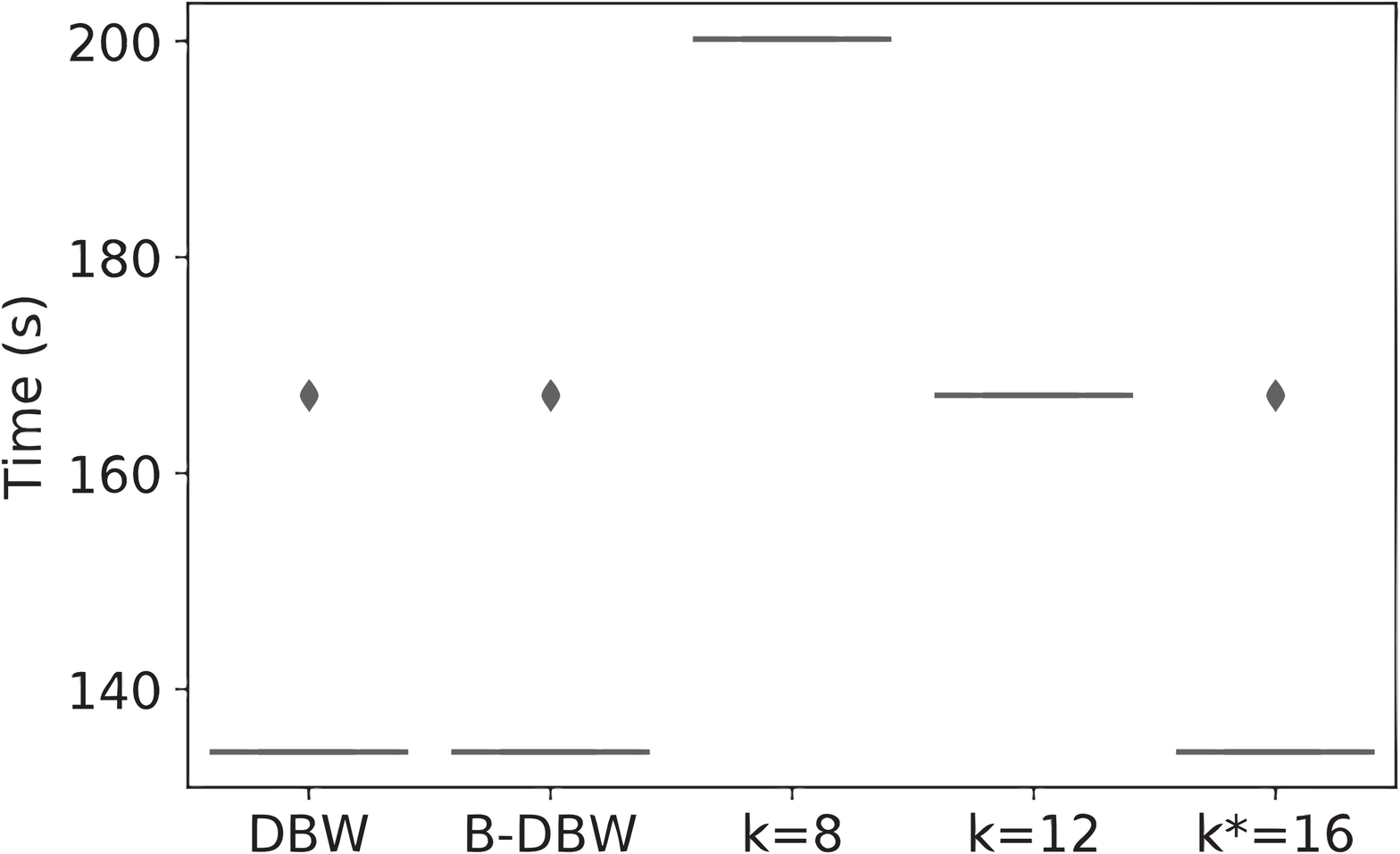}
                        \caption{\small  $\alpha = 0$}
        \end{subfigure}
        \hfill
        \begin{subfigure}[b]{0.32\textwidth}  
            \centering 
            \includegraphics[width=\textwidth]{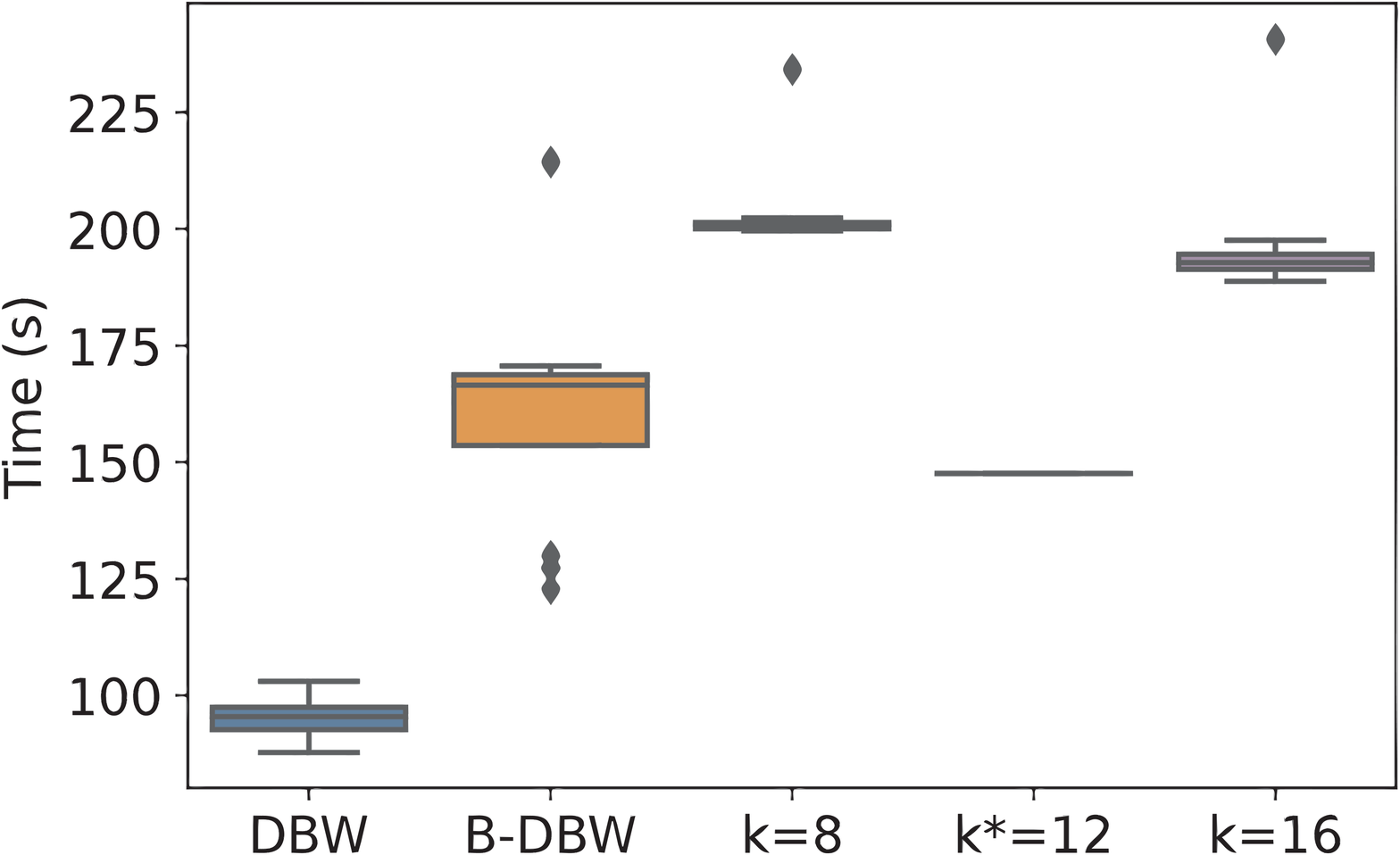}
                                    \caption{\small  $\alpha = 0.2$}
        \end{subfigure}
        %\vskip\baselineskip
        \hfill
        \begin{subfigure}[b]{0.32\textwidth}   
            \centering 
            \includegraphics[width=\textwidth]{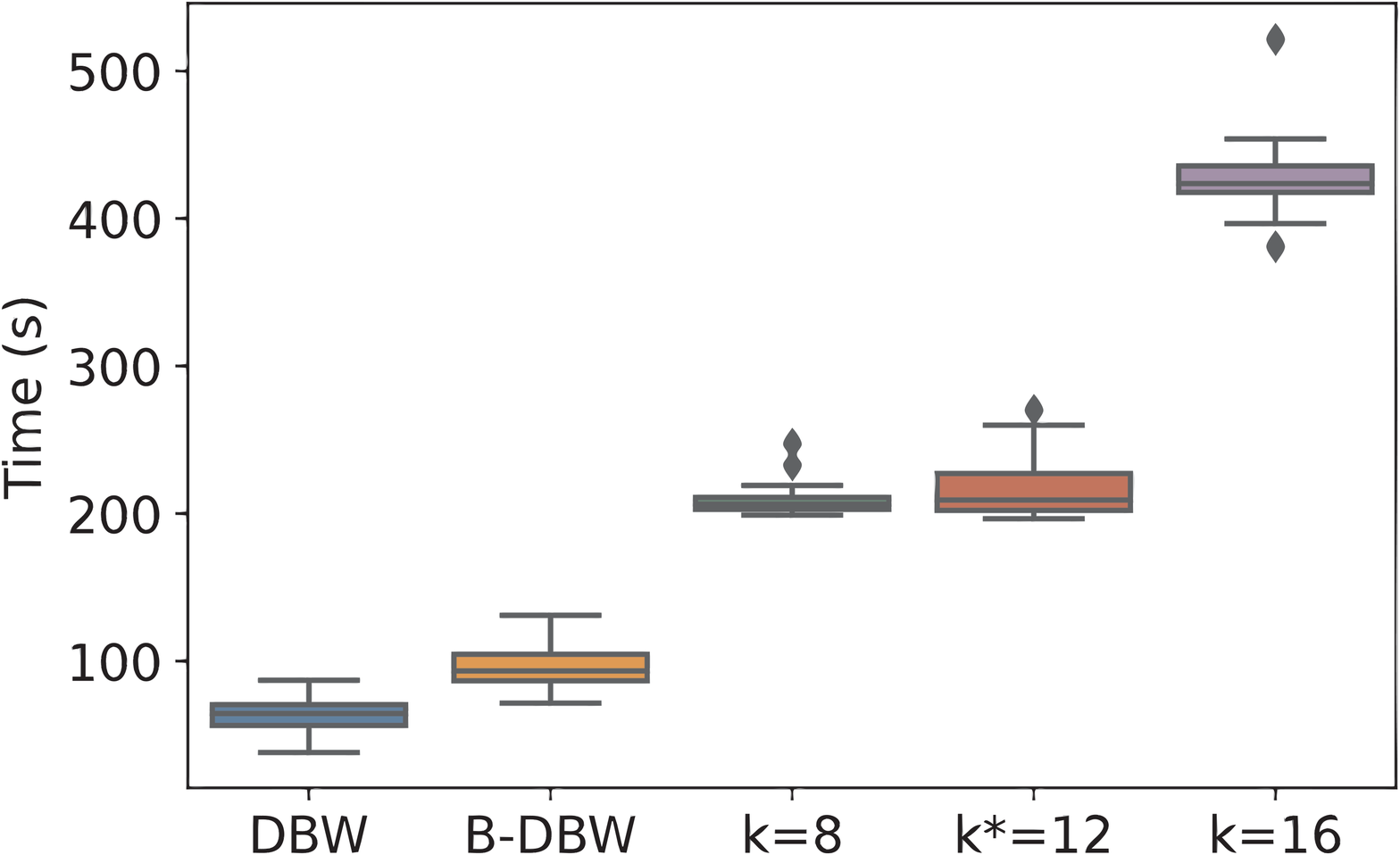}
                        \caption{\small  $\alpha = 1$}
                        \label{Fig:TimeDistributionC}
        \end{subfigure}
        
         \caption[ The average and standard deviation of critical parameters ]
        {\small Effect of round trip time distribution. MNIST, $n=16$ workers, batch size $B=500$, estimates computed over the last $D=5$ iterations, proportional rule for $\eta(k)$ in static settings where $\eta(k)=0.005k$.} 
        \label{Fig:TimeDistribution}
    \end{figure*}

Figures~\ref{fig:one_run_loss} and~\ref{fig:one_run_cifar_loss}  show, for a single run of the training process, the evolution of the loss over time and the corresponding choices of $k_t$ for the two dynamic algorithms. 
For static settings, the learning rate follows the proportional rule and the optimal static settings are $k^*=10$ for MNIST and $k^*=8$ for CIFAR10.
We can see that \DynamicBackUp{} achieves the fastest convergence across all other tested configurations of $k$, %(both static and through \bdbw), 
by using a different value of $k$ in different stages of the training process. 
In fact, as we have discussed after introducing~\eqref{e:exp_decrease}, the effect of $k$ on the gain depends on the module of the gradient and on the variability of the local gradients. In the bottom subplot, the dotted line shows how their ratio varies during the training process. For MNIST, up to iteration $38$, $\V(\vg_{i,t})$ is negligible in comparison to $\lVert \nabla F(\vw_t)\rVert^2$. \DynamicBackUp{} then selects small values for $k_t$ loosing a bit in terms of the gain, but significantly speeding up the duration of each iteration by only waiting for the fastest workers. As the parameter vector approaches a local minimum, $\lVert \nabla F(\vw_t)\rVert^2$ approaches zero, and the gain becomes more and more sensitive to $k$, so that \DynamicBackUp{} progressively increases $k_t$ up to reach $k_t=n=16$ as shown by the solid line. On the contrary \bdbw{} (the dashed line) selects most of the time $k_t=9$ with some variability to the randomness of the {estimates~$\widehat{T_{k,t}}$}.
For CIFAR10, as the stochastic gradients are more noisy, the ratio values $\lVert \nabla F(\vw_t)\rVert^2/\V(\vg_{i,t})$ are smaller than in MNIST, \DynamicBackUp{} selects higher values for $k_t$ (around 10) in the beginning of the training. After iteration $130$, the gain becomes more sensitive to $k$ and thus \DynamicBackUp{} progressively increases $k_t$ as observed in MNIST dataset. Note that \DynamicBackUp{} performs less advantageous in CIFAR10, although it is still the best one. As discussed in Sect.~\ref{sec:loss_decrease}, the gain~\eqref{e:exp_decrease} can be negative when the stochastic gradients are very noisy, which is the case for CIFAR10 dataset. This results in  \DynamicBackUp{} cautiously selecting $k_t=n$ according to~\eqref{e:greedy}, while the optimal $k_t$ at the iteration $t$ may be smaller. Note that working with significantly larger batch sizes would reduce the variability of the stochastic gradients.

Figures~\ref{fig:one_run_acc} and~\ref{fig:one_run_cifar_acc}  show, for a single run of the training process, the evolution of the test accuracy over time. We can see that \DynamicBackUp{} converges to a better model faster than the other methods for MNIST. The advantages of \DynamicBackUp{} on CIFAR10 are less evident on this specific run, but Figs.~\ref{fig:one_run_cifar_time_acc} and~\ref{fig:one_run_cifar_acc_time} show the distribution of the time to reach $80\%$ test accuracy and the distribution of the test accuracy after $200$ seconds using box plots.\footnote{The box shows the quartiles of the dataset while the whiskers extend to show the rest of the distribution. The middle bar gives the median value.}
On average \DynamicBackUp{} performs better than \bdbw{} or the optimal static setting.

\subsection{Round trip time effect}
In this subsection we consider round trip times (see Sect.~\ref{sec:time_estimation})
%\footnote{
%    Remember that round trip time includes the time to transmit the parameter vector from the PS to the worker, the time to compute the gradient, and the time to transfer the gradient to the PS.
%} 
are i.i.d.~according to a shifted exponential random variable $1-\alpha +\alpha \times \textrm{Exp}(1)$, where $0\le \alpha \le 1$.
%, as the assumption made in. 
We consider later realistic time distributions. 
This choice, common to~\cite{Lee, DuttaJGDN18}, allows us to easily tune the variability of the round trip times by changing $\alpha$. When $\alpha=0$, all gradients arrive at the same time at the PS, so that the PS should always aggregate all of them. As $\alpha$ changes from $0$ to $1$, the variance of the round trip times increases, and waiting for $k<n$ gradients becomes advantageous. 
%Here the learning rate is set by proportional rule for static settings. 
%In this first set of experiments we have followed set the learning rates according to the proportional rule with $\eta = 0.005k$. 

Figure~\ref{Fig:TimeDistribution} compares the time needed to reach a training loss smaller than $0.2$ for the two dynamic algorithms and the  static settings $k=16$, $k=12$, and $k=8$, that are optimal respectively for $\alpha = 0$, $\alpha=0.2$, and $\alpha=1$. 
For each of them, we carried out $20$ independent runs with different seeds.   
We find that our dynamic algorithm achieves the fastest convergence in all three scenarios, it is even 
$1.2$x faster and $3$x 
faster than the optimal static settings for $\alpha =0.2$ and $\alpha = 1$. 
There are two factors that determine this observation. 
First, as discussed for Fig.~\ref{fig:one_run}, there is no unique optimal value of $k$ to be used across the whole training process, and \DynamicBackUp{} manages to select the most indicated value in different stages of the training process. 
Second, \DynamicBackUp{} takes advantage of a larger learning rate. 
Both factors play a role. 
For example if we focus on Fig.~\ref{Fig:TimeDistributionC}, the learning rate for \DynamicBackUp{} is twice faster than that for $k=8$, but \DynamicBackUp{} is on average $3$x faster. Then, adapting $k$ achieves an additional $1.5$x improvement.
The importance of capturing the dynamics of the optimization process is again also evident by comparing \DynamicBackUp{} with \bdbw{}.  
While \bdbw{} takes advantage of a higher learning rate as well, it performs worse than our solution \DynamicBackUp{}.

%After running experiments of static methods on every $k$, the best static $k^*$ observed is $8$, $12$ and $16$ (w.r.t $\alpha = 0$, $0.8$ and $1$), as shown in 
%For the dynamic methods, we evaluate our $\DynamicBackUp$ (DBU) and another greedy one (k/T) similar to [Bayesian] that the numerator in \eqref{e:greedy} is changed to $k$. For both dynamic methods, the learning rates are set to $0.8$ which is the learning rate for the static $k=16$.

%In Fig.~\ref{Fig:TimeDistribution}, the y-axis presents the earliest time in seconds when the training loss decreases to $0.2$. 

%\vspace{-0.06cm}
\subsection{Batch size effect}
The batch size $B$ is another important hyper-parameter. It is often limited by the memory available at each worker, but can also be determined by generalization performance of the final model~\cite{hoffer17}.
In this subsection we highlight how $B$ also affects the optimal setting for $k$. These findings confirm that configuring the number of backup workers is indeed a difficult task, and knowing the characteristics of the underlying cluster is not sufficient. 

 \begin{figure}
     \centering
     \includegraphics[scale=0.25]{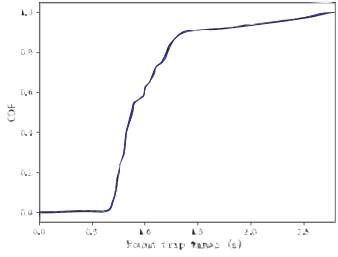}
     \caption{Empirical distribution of round trip times on a Spark cluster}
     \label{fig:cdf}
 \end{figure}
The experiments differ in two additional aspects from those in Fig.~\ref{Fig:TimeDistribution}. First, the distribution of the round trip times (shown in Fig.~\ref{fig:cdf}) is taken from a training a ML model through stochastic gradient descent on a production Spark cluster with sixteen servers, each with two 8-core Intel E5-2630 CPUs running at 2.40GHz. The cluster was managed using Zoe Analytics~\citep{pace17}.
%(the distribution is similar to~\cite[Fig.~7]{Lee}). 
Second, learning rates are configured according to the knee rule. We observe that the knee rule leads to a weaker variability of the learning rate in comparison to the proportional rule: for example, for $B=16$, $\eta$ increases by less than a factor $5$ when $k$ changes from $k=1$ to $k=16$, and it increases much less for larger $B$. 

 \begin{figure*}[t]
%\vspace{.3in}
       % \centering
        \hfill
        \begin{subfigure}[b]{0.32\textwidth}   
            \centering 
            \includegraphics[width=\textwidth]{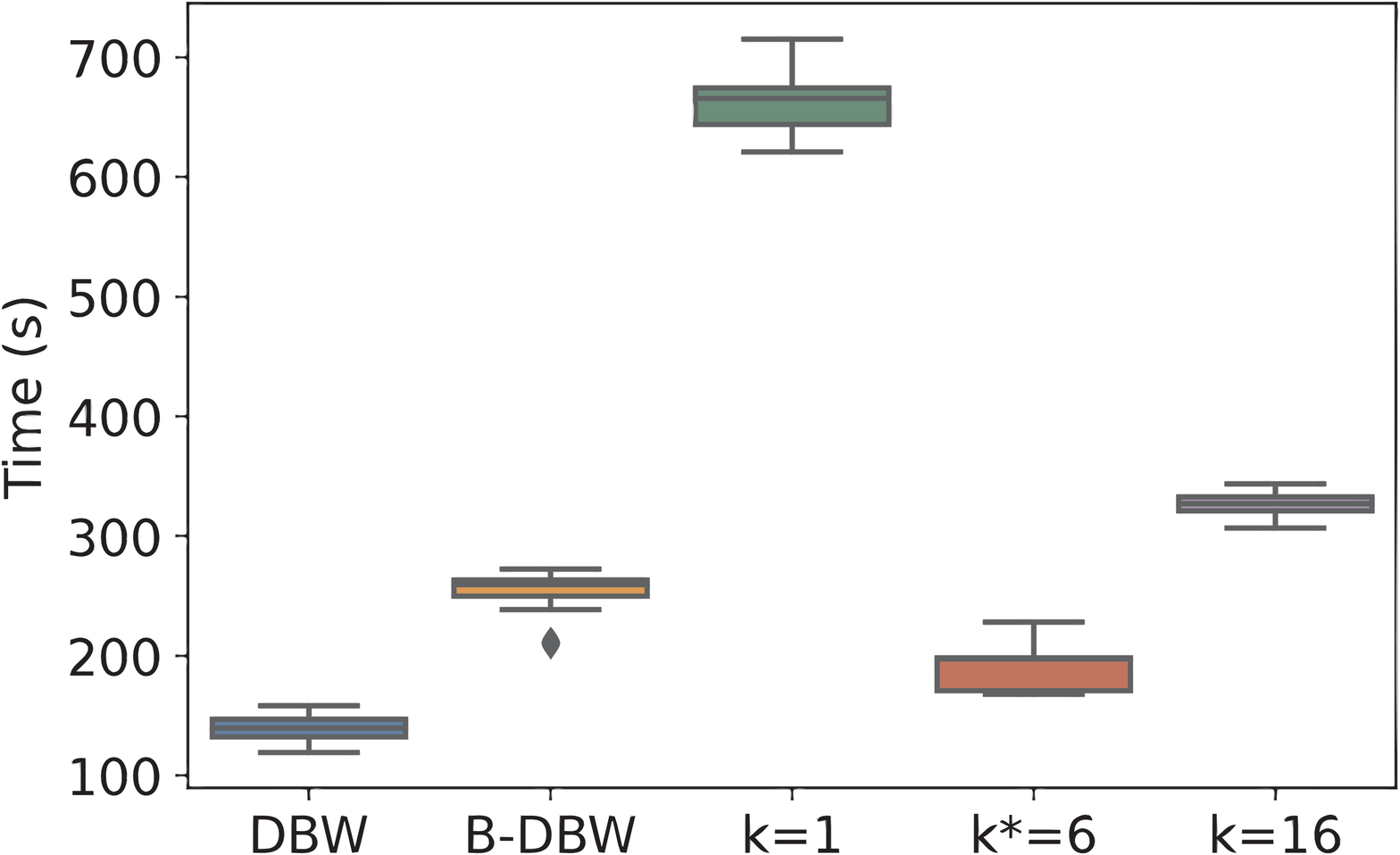}
            \caption{\small $B=16$, $\eta \in \{0.01,0.045,0.05\}$}
        \end{subfigure}
             \hfill   
                \begin{subfigure}[b]{0.32\textwidth}  
            \centering 
            \includegraphics[width=\textwidth]{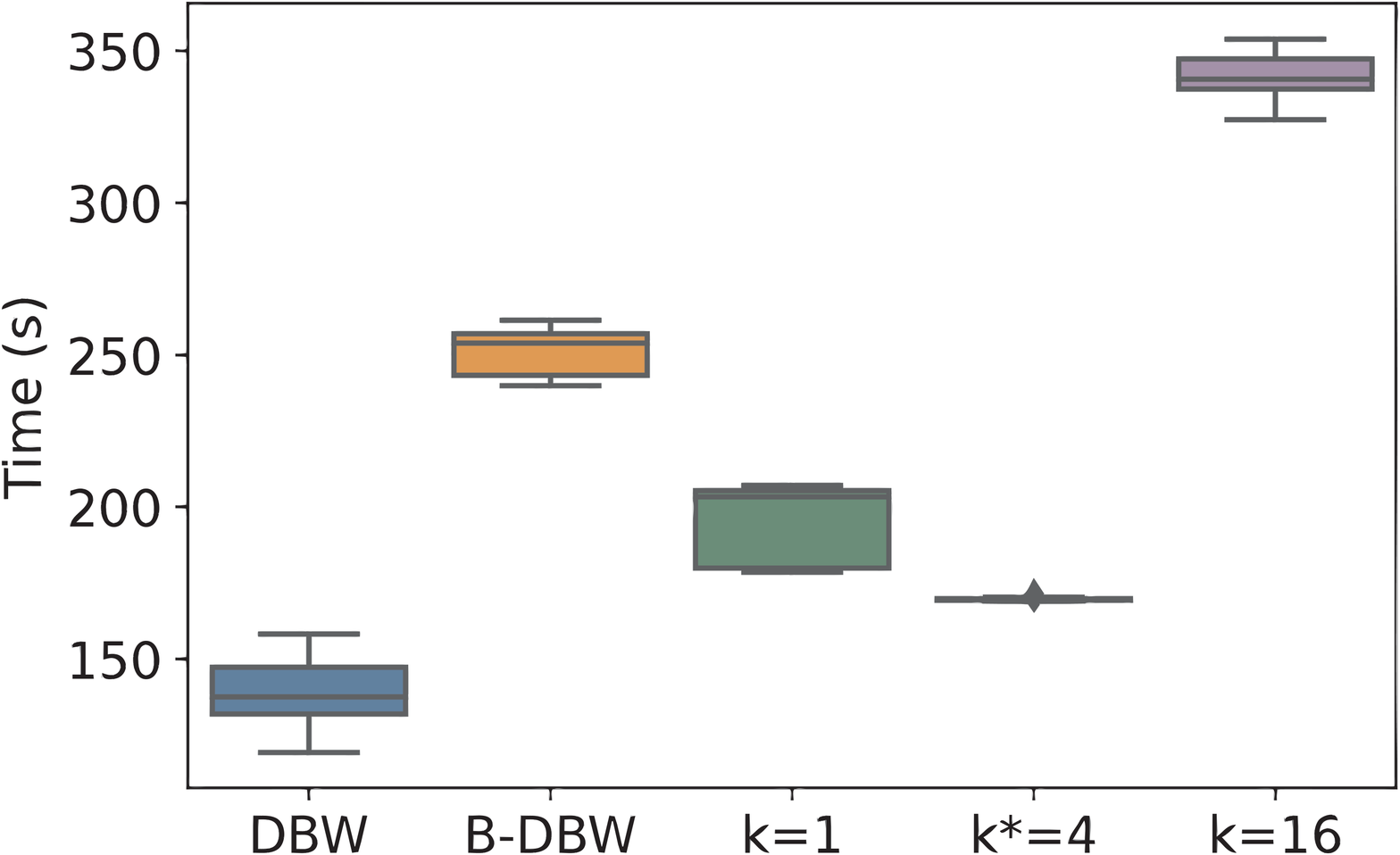}
                        \caption{\small $B=128$, $\eta \in \{0.04,0.044,0.05\}$}
        \end{subfigure}
        %\vskip\baselineskip
        \hfill
              \begin{subfigure}[b]{0.32\textwidth}
            \centering
            \includegraphics[width=\textwidth]{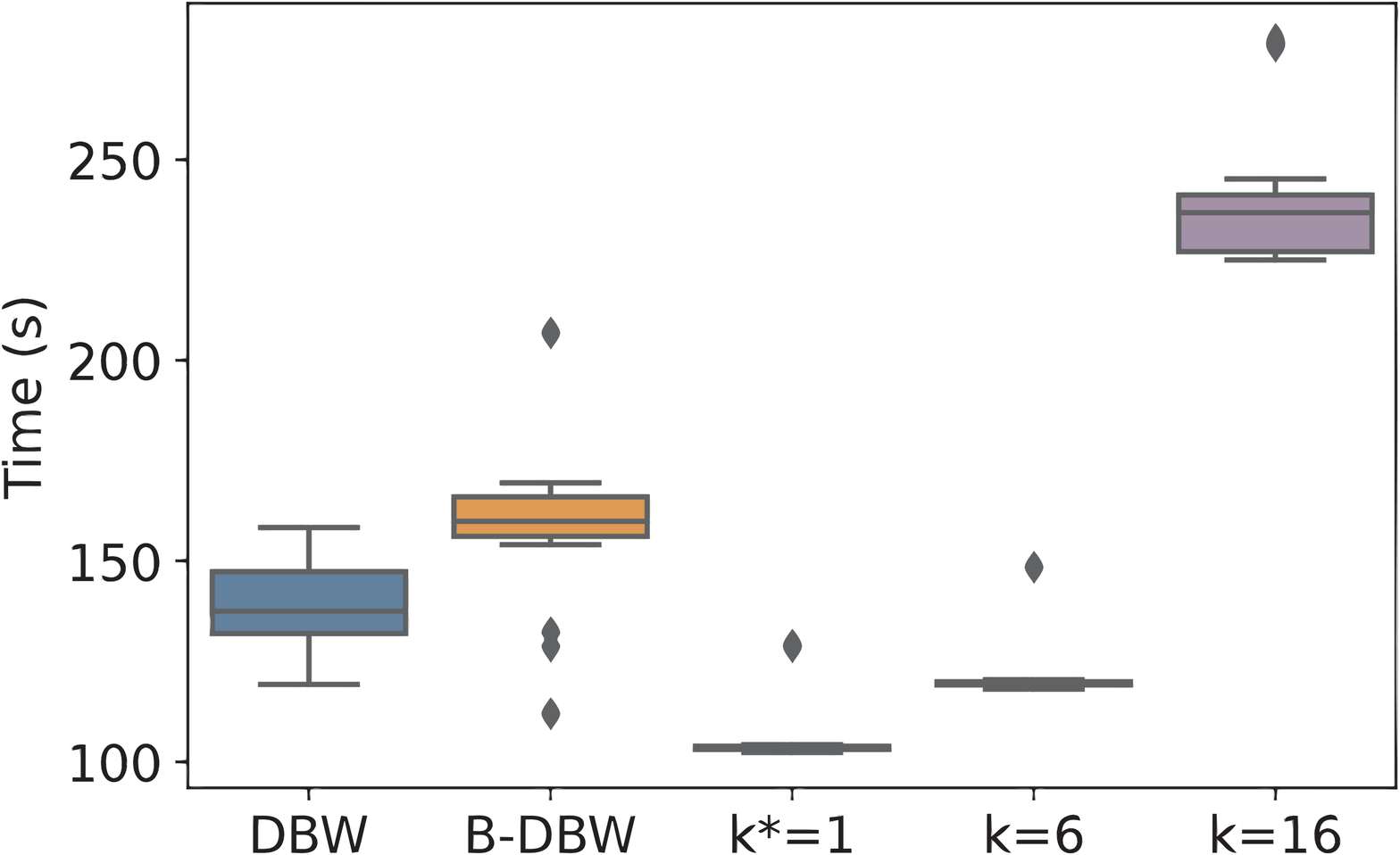}
            \caption{\small $B=500$, $\eta \in \{0.07,0.07,0.08\}$}
        \end{subfigure}
         \caption[ The average and standard deviation of critical parameters ]
        {\small Effect of batch size B. MNIST, $n=16$ workers, estimates computed over the last $D=5$ iterations, knee rule for $\eta$ in static settings with values shown above for each $k$.} 
        \label{Fig:BatchSize}
    \end{figure*}

%As a consequence, we expect the improvements of \DynamicBackUp{} or \bdbw{} in comparison to static settings with a small value of $k$ to be reduced.

Figure~\ref{Fig:BatchSize} shows the results for $B=16, 128, 500$, comparing the dynamic methods with a few static settings, including the optimal static one that decreases from $k^*=6$ for $B=16$ to $k^*=1$ for $B=500$. 
Again, Equation~\eqref{e:exp_decrease} helps to understand this change of the optimal static setting with different batch size: as the batch size increases, the variability of gradients decreases, so that the numerator depends less on $k$. The advantage of reducing $T_{k,t}$ by selecting a small $k$ can compensate the corresponding decrease of the gain $\mathcal G_{k,t}$.

Since learning rates chosen by the knee rule for the static settings are now close to dynamic ones, \DynamicBackUp{} does not outperform the optimal static setting, but its performance are quite close, and significantly better than \bdbw{} for $B=128, 500$. It is worthy to stress that, when running a given ML problem on a specific cluster environment, the user cannot predict the optimal static setting $k^*$ without running preliminary short training experiments for every $k$. \DynamicBackUp{} does not need them.
%Moreover, the \bdbw{} scheme is not as robust as the proposed \DynamicBackUp{}, since it chooses $k$ independently from the batch size. 

\begin{figure}
    \centering
    \includegraphics[scale=0.05]{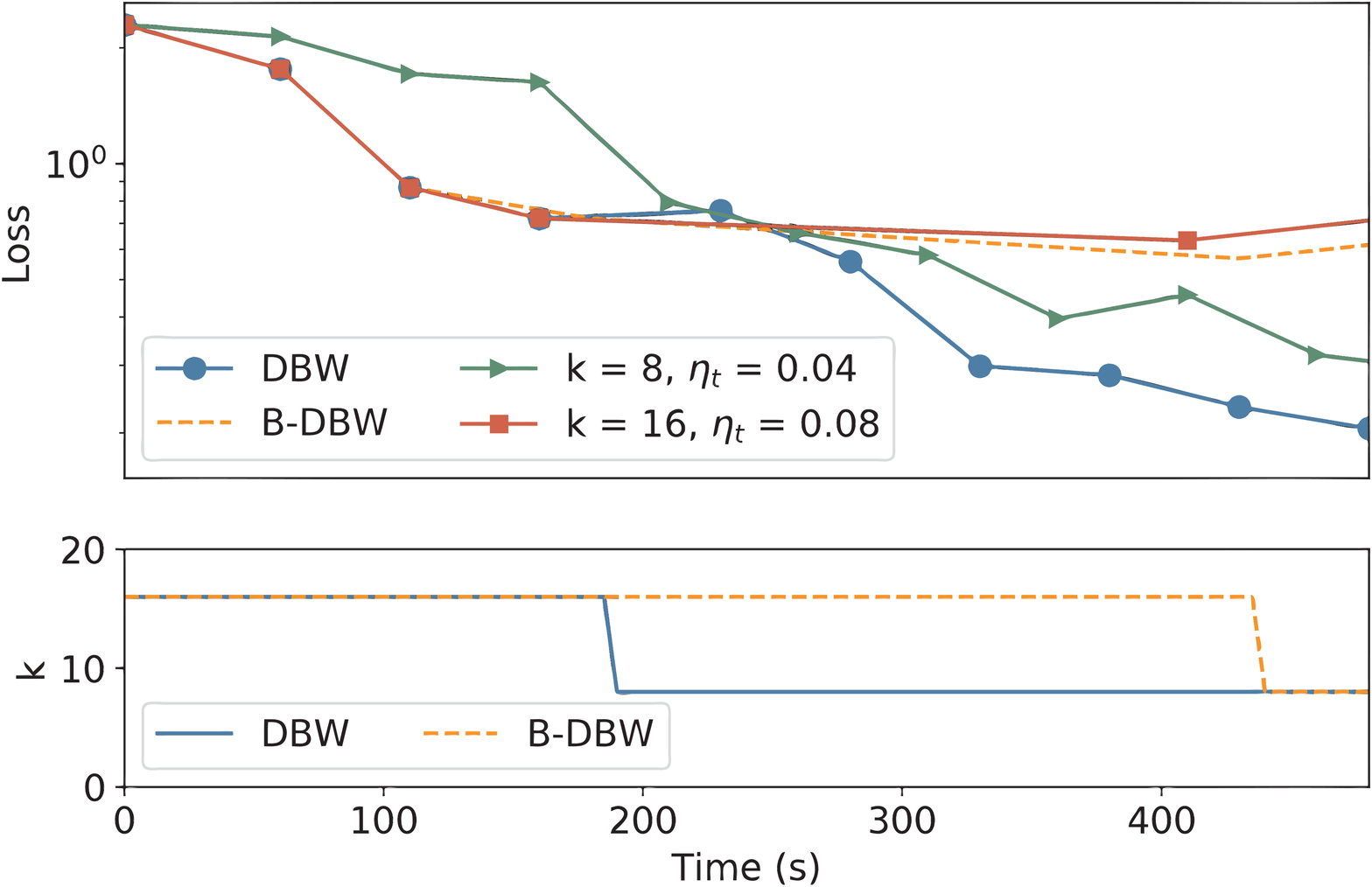}
    \caption{Robustness to slowdowns of the system. MNIST, $n=16$ workers, batch size $B=500$, estimates computed over the last $D=5$ iterations, proportional rule for $\eta(k)$ in static settings where $\eta(k)=0.005k$.}
    \label{fig:dynamic}
\end{figure}

\subsection{Robustness to slowdowns}
\label{subsec:robust}
Until now, we have considered a stationary setting where the distribution of round trip times does not change during the training.
Figure~\ref{fig:dynamic} shows an experiment in which half of the workers experience a sudden slowdown during the training process. Initially, round trip times are all equal and deterministic, so that the optimal setting is $k_t=n=16$. Suddenly, at time $t=160$s, half of the workers in the clusters slow down by a factor 5 and the optimal static configuration is now to select $k_t=n/2=8$.  We can see that \DynamicBackUp{} detects the slowdowns in the system and then correctly selects $k_t=8$.

\subsection{Comparison with \adasync}
\label{subsec:adasync}
\begin{figure*}
    \hfill
    \begin{subfigure}[b]{0.58\textwidth}
            \centering
            \includegraphics[width=\textwidth]{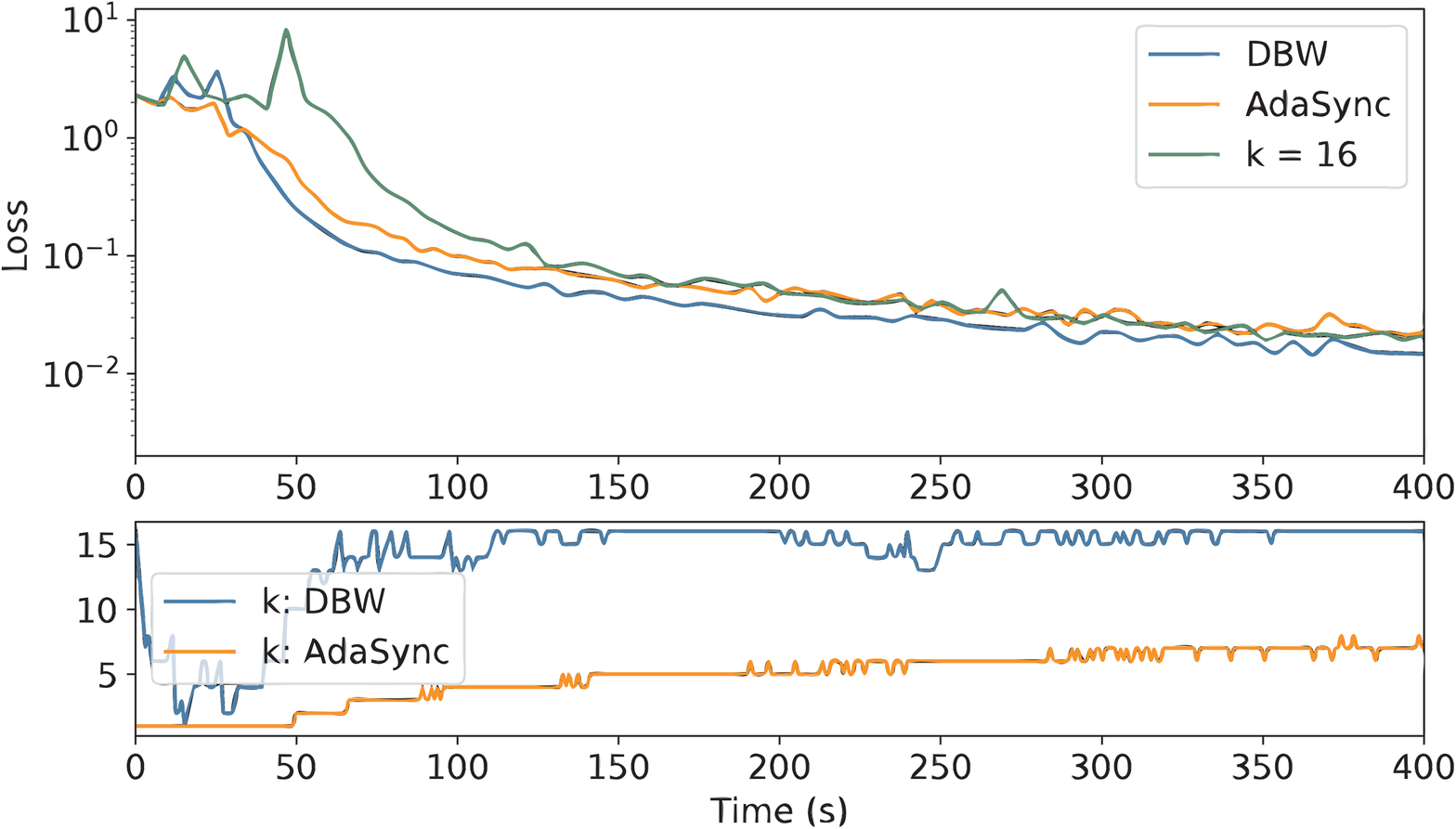}
            \caption{Loss versus Time ($\alpha = 0.1$)}
            \label{fig:one_run_adasync}
    \end{subfigure}
    \begin{subfigure}[b]{0.4\textwidth}
            \centering
            \includegraphics[width=\textwidth]{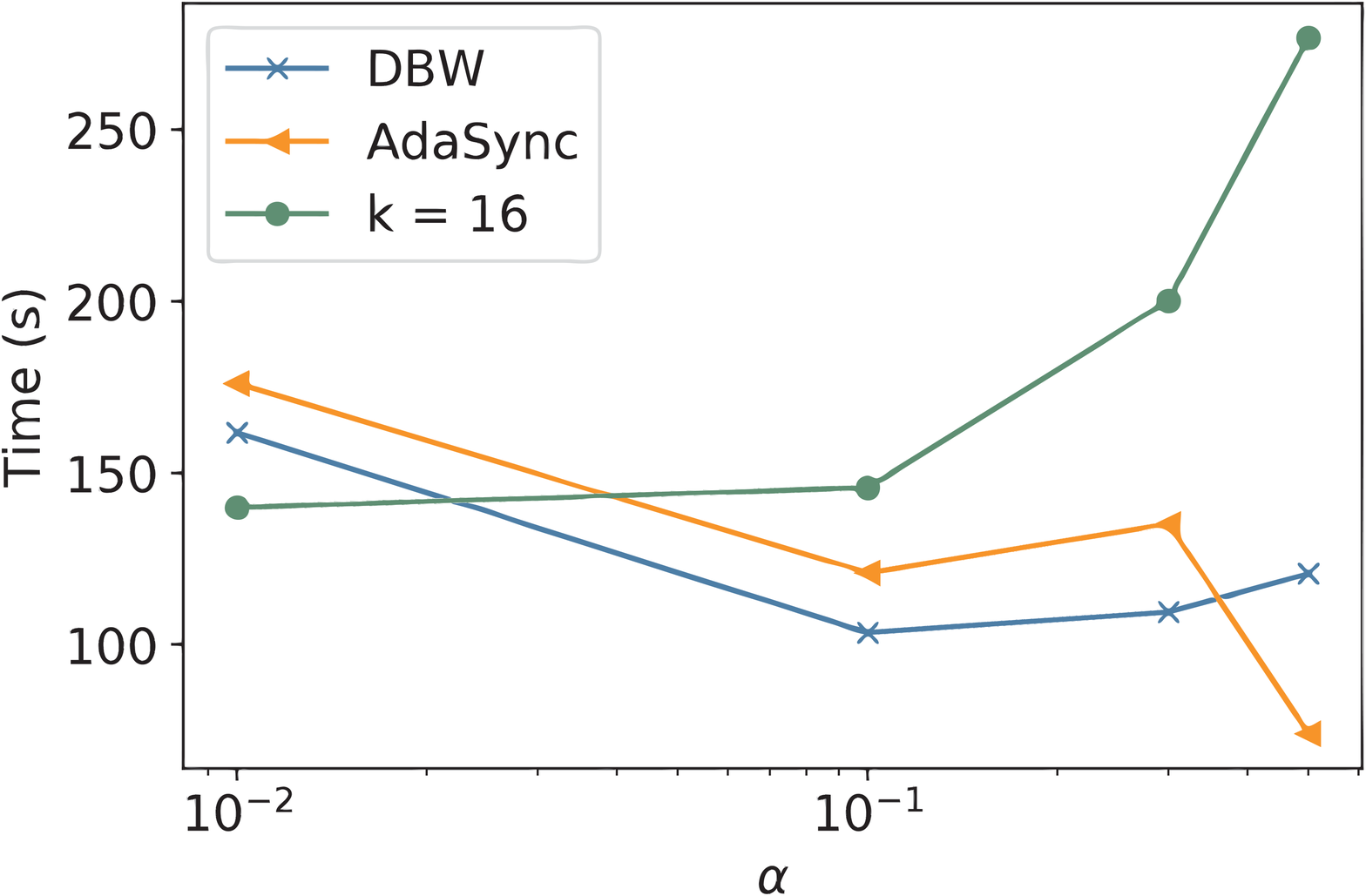}
            \caption{Average convergence time versus different $\alpha$}
            \label{fig:vary_alpha}
    \end{subfigure}
    \caption{Training on MNIST, batch size $B = 500$, $n = 16$ workers, estimates computed over the last $D = 5$ iterations. $\eta = 0.08$. Round trip times follow shifted exponential distribution $1-\alpha+\alpha \textrm{Exp}(1)$}
\end{figure*}

\adasync{}~\cite{dutta} is a dynamic backup scheme designed for the Push and Interrupt (PsI) case, under the assumption that the round trip times follow shifted exponential distribution. For the comparison, we consider then this setting.
%To fairly compare with, in this subsection, we consider PsI and the shifted exponential distribution, i.e.,~$1-\alpha+\alpha\times \textrm{Exp(1)}$. 
For \adasync, the quadratic formulation in~\cite[Appendix D.1]{dutta} is used to derive the number of backup workers.  \adasync{} updates  $k$ at the end of a time-window. We consider this time-window small enough  for \adasync{} evaluating the possibility to update $k_t$ at each iteration, as \DynamicBackUp{} does.

Figure~\ref{fig:one_run_adasync} shows, for a single run of the training process, the evolution of the loss over time and the corresponding choices of $k_t$ for \DynamicBackUp{} and \adasync, when $\alpha = 0.1$, i.e.,~round trip times follow distribution $0.9+0.1\textrm{Exp}(1)$. \DynamicBackUp{} quickly reaches a large value of $k_t$ close to $n$. For small $\alpha$ the variance of round trip time is small, so choosing large $k_t$ does not lead large iteration times $\mathbb
E[T_{k,t}]$ but benefits the gain in \eqref{e:exp_decrease}.
The approximated formula used by \adasync{}, even if derived under the assumption of shifted exponential distributions, does not depend on $\alpha$, and \adasync{} fails to increase fast the value of $k_t$. 
%but only 
%However, this effect of $\alpha$ is not considered in the dynamic scheme of \adasync{} as the quadratic formulation for choosing $k_t$ is $\alpha$ independent. Therefore, in the case of small $\alpha$, \adasync{} converges slower than $\DynamicBackUp{}$.

Fig.~\ref{fig:vary_alpha} shows 
the average convergence time\footnote{The convergence time noted here is the time when the training loss reaches 0.07.} computed over 10 independent runs under different $\alpha$. The larger $\alpha$, the larger the variance of round trip times.
We can see that when $\alpha$ is smaller than 0.3, \DynamicBackUp{} performs better than \adasync{}. While, \adasync{} works better for  larger $\alpha$, which suggests \DynamicBackUp{} may be too conservative on the number of backup workers in the late phase of the training. 
%There is still space for the improvement.

Remember that the estimated gain $\widehat{\mathcal G_{k,t}}$ used in \eqref{e:greedy} for choosing $k_t$, is a lower bound for the true loss decrease. In the late training phase, when the gradient norm becomes smaller, small values of $k$ may lead to estimate a negative (see~\eqref{e:gain}). In this case,
\DynamicBackUp{} conservatively chooses a larger $k$ for which the gain is estimated to be positive.
On the other hand, \adasync{} requires prior knowledge on the round trip time distribution. %, where the explicit formulation for $T_{k,t}$ is required. 
This distribution may be hard to estimate and may change during the training period, that is often very long for state-of-the-art machine learning models (e.g., weeks).
%However, this is not really practical, not only the time distribution information is hard to know, but also this information could be online dynamic as shown by one of the cases in Sect.~\ref{subsec:robust}. 
Notice that \DynamicBackUp{} does not require any prior knowledge on the system.

\section{Conclusions}
\label{s:conclusions}
In this paper, we have shown that the number of backup workers needs to be adapted at run-time and the correct choice is inextricably bounded, not only to the  cluster's configuration and workload, but also to the hyper-parameters of the learning algorithm and the stage of the training. 
We have proposed a simple algorithm  \DynamicBackUp{} that, without prior knowledge about the cluster or the problem, %and almost no overhead,
achieves good performance across a variety of scenarios, and even outperforms in some cases the optimal static setting.

As a future research direction, we want to 
%improve the performance of \DynamicBackUp{} (e.g.~by refining the estimate of the gain currently based on the lower-bound in~\eqref{e:lower_bound}) and to 
extend the scope of \DynamicBackUp{} to dynamic resource allocation, e.g.,~by automatically releasing computing resources if $k_t<n$ and the fastest $k_t$ gradients are always coming from the same set of workers.
In general, we believe that distributed systems for ML are in need of adaptive algorithms in the same spirit of the utility-based congestion control schemes  developed in our community starting from the seminal paper~\cite{kelly98}. As our work points out, it is important to define new utility functions that take into account the learning process.
Adaptive algorithms are even more needed in the federated learning scenario~\cite{konecny15}, where ML training is no more relegated to the cloud, but it occurs in the wild over the whole internet.
Our paper shows that even simple algorithms can provide significant improvements.

\section{Acknowledgements}
This work has been carried out in the framework of a common lab
agreement between Inria and Nokia Bell Labs (ADR 'Rethinking the Network'). We thank~Alain Jean-Marie for having suggested the estimation technique in Sect.~\ref{sec:time_estimation} and Pietro Michiardi for many helpful discussions. 

\appendix
\section{Proof of $\E[\mathcal{T}_{i,i}] \le \E[\mathcal{T}_{i+1,i+1}]$}
\label{a:proof}
Remember that we assume that $T_{k,t}$ depends on the past only through the number of workers $k_{t-1}$ selected at the previous iteration. This approximation is correct when round trip times are exponentially distributed. We start proving the inequality under the assumption that round trip times are exponentially distributed. We move then to the general case. 

Consider the beginning of a new iteration $t$ when the PS systematically waits for $i+1$ nodes. Without loss of generality, let us assume that the workers who finished the computation are labeled $1,2, \dots, i+1$. Worker $j\le i+1$  needs an exponentially distributed round trip time $\omega_j$ to complete the new computation. Worker $j > i+1$ needs to complete iteration $t-1$, with residual time $\omega_j'$, and possibly start a new one with the updated parameter vector, with corresponding residual time $\omega_j$; both $\omega_j$ and $\omega_j'$ are exponentially distributed.

Let $\mu(l,A)$ denote the $l$-th smallest element of the multiset $A$.
The duration of the new iteration is then  $T_{i+1,t} = \mu(i+1,\{\omega_1, \dots, \omega_i, \omega_{i+1}, \omega_{i+2}'+ \omega_{i+2}, \dots, \omega_{n}' + \omega_{n}\})$.

Now consider the case when the PS only waits for the $i$ workers. Again we assume the the first workers who finished the iteration are labeled $1,2, \dots, i$. We also couple all the round trip times so that $\omega_j$ for $j =1, \dots, n$ and $\omega'_j$ for $j=i+2, \dots, n$ denote the same quantities and have the same values. In this case also worker $i+1$ needs to terminate the previous computation; this will require a time $\omega_{i+1}'$, but its specific value is irrelevant.
The duration of the new iteration is  $T_{i,t} = \mu(i,\{\omega_1, \dots, \omega_i, \omega_{i+1}'+ \omega_{i+1}, \omega_{i+2}'+ \omega_{i+2}, \dots, \omega_{n}' + \omega_{n}\})$.

\begin{align*}
    T_{i+1,t} &  = \mu\!\left(i+1,\{\omega_1, \dots, \omega_i, \omega_{i+1}, \omega_{i+2}'+ \omega_{i+2}, \dots, \omega_{n}' + \omega_{n}\}\right)\\
        & \ge \mu\!\left(i+1,\{\omega_1, \dots, \omega_i, 0, \omega_{i+2}'+ \omega_{i+2}, \dots, \omega_{n}' + \omega_{n}\}\right)\\
        & =\mu\!\left(i,\{\omega_1, \dots, \omega_i,  \omega_{i+2}'+ \omega_{i+2}, \dots, \omega_{n}' + \omega_{n}\}\right)\\
        & \ge \mu\!\left(i,\{\omega_1, \dots, \omega_i, \omega_{i+1}'+ \omega_{i+1}, \omega_{i+2}'+ \omega_{i+2}, \dots, \omega_{n}' + \omega_{n}\}\right)\\
        & = T_{i,t},
\end{align*}
where the first inequality follows from the fact that replacing an element in the set with a smaller one can only decrease the $(i+1)$-th smallest element of the multiset, the second equality from the fact that $0$ is necessarily the smallest value in the multiset, and the last inequality from the fact that enlarging a multiset cannot increase its $i$-th smallest element.

In the general case, we show that the time at which the $t$-th iteration will start is not larger when the PS waits for $i$ workers than when it waits for $i+1$ workers. We will couple the round trip times so that in both cases the duration of the $m$-th round trip time for worker $j$ is the same in both systems. % and we denote it as $\psi_{j,m}$. 

Let $\chi_{i,t}$ denote the time at which the $t$-th system iteration starts when then PS waits for $i$ workers. We also consider a \emph{lazy} system, where the PS does not need to start the new iteration as soon as $i$ new updates are available, but it can start after an arbitrary delay. We say that  a sequence $(\chi^{(l)}_{i,t})_{t\in \mathbb N}$ is feasible for the lazy system, if it corresponds to a valid sequence of starting times. We observe that for any feasible sequence $\chi^{(l)}_{i,t} \ge \chi_{i,t}$ for each $t$ as the lazy system can only introduce slack times. Finally, we note that $(\chi_{i+1,t})_{t\in \mathbb N}$ is a feasible sequence for the lazy system, as at each time $\chi_{i+1,t}$, the system has available $i$ new updates (it has $i+1$) and can then start a new iteration. It follows that $\chi_{i+1,t} \ge \chi_{i,t}$.

\bibliographystyle{elsarticle-num}
\bibliography{ref}

\end{document}